\documentclass[9pt,twocolumn,twoside]{pnas-new}

\templatetype{pnasresearcharticle} 

\title{Discovering Sensorimotor Agency in Cellular Automata using Diversity Search}

\author[a,1,2]{Gautier Hamon}
\author[a,b,1,2]{Mayalen Etcheverry} 
\author[a,c,2]{Bert Wang-Chak Chan}
\author[a,2]{Clément Moulin-Frier}
\author[a,2]{Pierre-Yves Oudeyer}
\affil[a]{INRIA, University of Bordeaux, Talence 33405, France}
\affil[b]{Poietis, Pessac 33600, France}
\affil[c]{Google DeepMind, Tokyo, Japan}

\leadauthor{Hamon} 

\significancestatement{In the quest to understand the origins of life, a key mystery is understanding
which environmental conditions can give rise to self-organization of ``individuals'',
starting from low-level uncoordinated molecular components. Cellular automata
have long been used as theoretical models of morphogenesis, but even in these simulated
worlds, understanding how to generate self-organizing robust ``individuals'' has
remained an open question. In this paper, we leverage recent advances 
in machine learning, including diversity search algorithms, to show how one can
systematically find environmental conditions in CA leading to self-organization of basic
forms of agency. The discovered agents show surprisingly robust capabilities to
move, maintain their body integrity and navigate among various obstacles.
This approach opens new perspectives in AI and synthetic bioengineering.}

\authorcontributions{G.H.,M.E.,B.C.,C.MF. and PY.O. designed research; G.H. and M.E. performed  research;  G.H. and M.E. analyzed data; and G.H.,M.E.,B.C.,C.MF. and PY.O. wrote the paper.}
\equalauthors{\textsuperscript{1}A.O.(Author One) contributed equally to this work with A.T. (Author Two) .}
\correspondingauthor{\textsuperscript{2}To whom correspondence should be addressed. E-mail: \{gautier.hamon, mayalen.etcheverry, clement.moulin-frier, pierre-yves.oudeyer\}@inria.fr,  bertchan@google.com}


\begin{abstract}
The research field of Artificial Life studies how life-like phenomena such as autopoiesis, agency, or self-regulation can self-organize in computer simulations. In cellular automata (CA), a key open-question has been whether it it is possible to find environment rules that self-organize robust “individuals” from an initial state with no prior existence of things like “bodies”, “brain”, “perception” or “action”. In this paper, we leverage recent advances in machine learning, combining algorithms for diversity search, curriculum learning and gradient descent, to automate the search of such “individuals”, i.e. localized structures that move around with the ability to react in a coherent manner to external obstacles and maintain their integrity, hence primitive forms of sensorimotor agency.  We show that this approach enables to find systematically environmental conditions in CA leading to self-organization of such basic forms of agency. Through multiple experiments, we show that the discovered agents have surprisingly robust capabilities to move, maintain their body integrity and navigate among various obstacles. They also show strong generalization abilities, with robustness to changes of scale, random updates or perturbations from the environment not seen during training. We discuss how this approach opens new perspectives in AI and synthetic bioengineering.
\end{abstract}

\begin{document}

\maketitle

\ifthenelse{\boolean{shortarticle}}{\ifthenelse{\boolean{singlecolumn}}{\abscontentformatted}{\abscontent}}{}

\begin{figure*}[t!]
\centering
\includegraphics[width=0.9\textwidth]{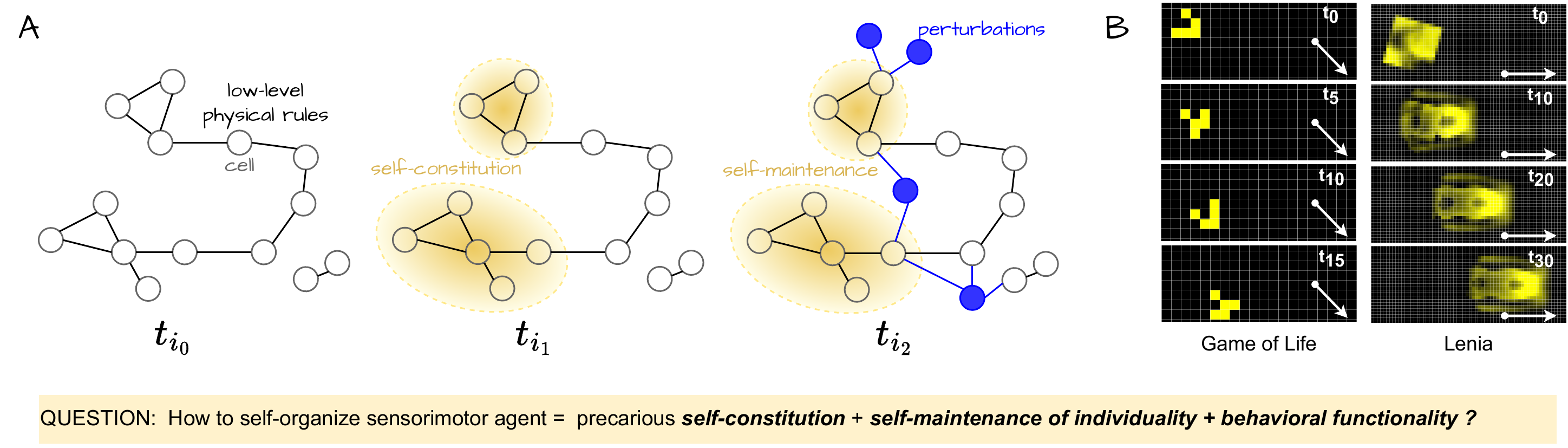}
\caption{\textbf{Overview of the scientific question}. (A) The enactivist framework: ($t_{i_0}$) In the beginning there is only an environment made of low-level elements (cells) and physical laws (local rules). There is no prior notion of agency, no body, no sensor. ($t_{i_1}$) Agents can come to existence through the coordination of the low-level elements (self-constitution of individuality). ($t_{i_2}$) To maintain their integrity, agents must sense and react to perturbations using only local update rules (self-maintenance of individuality). (B) In cellular automata models like the Game of Life and a more complex continuous extension called Lenia, it was shown that it is possible to self-organize so-called \textit{gliders} i.e. spatially-localized patterns with directional movement. Directional movement (white arrows) and timesteps are displayed. (Question) In this work, following the enactivist modeling framework, we try to answer the following scientific question: is it possible to find environments in which a subpart could self-organize and be called a ``sensorimotor agent''? This would require the existence and emergence of gliders-like structures that not only self-constitute and show motility, but that are also robust to external perturbations and hence must develop some form of sensorimotor apparatus enabling them to make ``decision'' and ``sense'' at the macro scale through local interactions only.}
\label{fig:overview}
\end{figure*}

\dropcap{U}nderstanding what has led to the emergence of life, cognition and natural agency as we observe in living organisms has been a central debate across many sectors of life sciences. Biological organisms are made of collections of cells that follow low-level distributed rules and yet they constitute a coherent unitary whole, displaying strong \textit{individuality}\footnote{ability of a self-organizing structure (subpart of the environment) to preserve and propagate some spatiotemporal unity \cite{krakauer2020information}, making it a distinguishable coherent entity in the domain in which it exists} and \textit{self-maintenance}\footnote{ability of a self-organizing structure to modify its interactions with the rest of the environment for maintaining its integrity} in their environment, what was described to be an \textit{autopoietic system}\footnote{Introduced by Maturana and Varela~\cite{maturana1980autopoiesis}, the concept of autopoiesis refers to a system capable of producing and maintaining itself by creating its own parts}. While a central concept in theoretical biology, the characterization of an autopoietic system and the understanding of the processes underlying its self-organization remain a live issue. Further demystifying how those processes do not just give rise to organic individuation\footnote{regulation at the metabolic, transcriptional and morphological level to maintain organic integrity~\cite{di2019process}} but also to sensorimotor\footnote{active engagement in loops of actions and perceptions in the external environment~\cite{di2019process}} and even intersubjective\footnote{active engagement in communicative interactions and structural coupling with other agents~\cite{di2019process}} agency, is at the center of the debate~\cite{di2019process}. In fact, recent advances in biology and basal cognition suggest that many autopoietic systems that we find in nature, including plants and brainless animals, are robust sensorimotor agents capable of using a body for sensing opportunities, computing decisions and acting in their environment~\cite{lyon2021reframing}. 
The pragmatic and complementary question to the debate, central in artificial life (ALife) and artificial intelligence (AI) research, is: can we engineer the necessary ingredients leading to the emergence of functional forms of life and sensorimotor agency in an artificial substrata in which initially there is literally no body (and thus no sensing, no acting, no agent)? Although there is already a large body of work that proposes to study the emergence of life and cognition in agents-as-they-could-be, it is generally done either by jumping over the biological processes that enable organisms to survive (the \textit{mechanistic view}, as in e.g. reinforcement learning, which considers a pre-existing agent with predefined sensors and actuators) or inconclusive so-far in showcasing higher-level forms of sensorimotor agency (the \textit{enactivist view}, as in e.g. artificial chemistry which studies how some form of agency can emerge from low-level chemical reactions). Herein, after giving some background on the mechanistic and enactivist views on cognition and on their respective limitations, we suggest that modern tools from machine learning (ML) can help us bridge the gap between those two views. Whereas those tools have mainly been deployed within the mechanistic framework, we show that they can efficiently assist the discovery of environments that self-organize relatively-advanced forms of sensorimotor agency whose existence and understanding is fundamental within the enactivist framework for supporting theories about the origins of life and cognition.  

In the mechanistic view, one assumes the existence of agents that have well defined physical body and information processing brain allowing them to interact with the rest of the environment through predefined sensors and actuators. Robots for instance are referred as embodied agents: their individuality is clear, as they can easily be distinguished from the rest of the environment, and their self-maintenance is often not a problem, as their body does not change over time except for rare cases of real world or artificially-induced degradation. Hence it is not questioned what makes an agent an agent or even what makes a body a body~\cite{di2019process}. Rather, a more central question is to understand how higher-level cognitive processes and sensorimotor adaptivity can arise in the agent through its interactions with the environment. A common methodology is the generation of a distribution of environments (tasks and rewards) and the use of learning approaches, such as deep reinforcement learning, to train the agent’s brain to master and generalize those tasks. Within that framework, it was shown that it is possible to engineer agents capable of repertoires of advanced sensorimotor skills such as precise locomotion~\cite{DBLP:journals/corr/abs-1901-01753}, object manipulation~\cite{akkaya2019solving}, tool use~\cite{baker2019emergent} and even capable of adapting the learned behaviors to unseen environmental conditions~\cite{team2021open}. Interestingly, they show that the use of curriculum learning\footnote{family of mechanisms that adapt the distribution of training environments to the learner capabilities} is crucial to generate generally capable agents.
However, the clear body/brain/environment distinction of the mechanistic framework bears little resemblance with the way information seems to be processed by biological systems. Notably it goes against the concept of morphological computation~\cite{pfeifer2006body}, which argues that all physical processes of the body, not only electrical circuitry in the brain but also morphological growth and body reconfiguration, are integral parts of cognition and can achieve advanced forms of computation.

The enactive view on embodiment however is rooted in the bottom-up organizational principles of living organisms in the biological world. The modeling framework typically uses tools from dynamical and complex systems theory where an artificial system (the environment) is made of low-level elements of matter (called atoms, molecules or cells) described by their inner states (e.g. energy level) and locally interacting via physics-like rules (flow of matter and energy within the elements) (Fig.\ref{fig:overview}A-$t_{i_0}$). There is no predefined notion of agent embodiment, instead it is considered that the body of the agent must come to existence through the coordination of the low-level elements (Fig.\ref{fig:overview}A-$t_{i_1}$) and must operate under environmental perturbations and precarious conditions\footnote{the idea that bodies are constantly subjected to disruptions and breakdowns~\cite{di2019process}} (Fig.\ref{fig:overview}A-$t_{i_2}$). 
Hence, the self-constitution and self-maintenance of individuality are prior conditions for any agency to emerge as it determines the agent’s own existence and survival \cite{di2019process}. 
This shifts the problem of ``building agents as-they-could-be'' to a problem of engineering second-order emergence~\cite{froese2009enactive}: how to design environments that can give rise to self-constituting agents that, coupled with the rest of environment, give rise to sensorimotor behaviors?
Previous work has shown that the realisation of autopoietic entities in computational media is possible~\cite{VARELA1974187,mcmullin2004thirty,beer2004autopoiesis, agmon2015ontogeny}. For instance, fully emergent structures showing spatial localization and movement have been discovered, such as the well-known gliders in the game of life up to richer life-like patterns in continuous models of cellular automata (Fig.\ref{fig:overview}B). So far however, two major challenges remain poorly addressed in the enactivist literature. First, autopoietic structures have so far mainly been discovered by human eye and as the result of time-consuming manual search, limiting their discovery and analysis. While some recent works, based on information theory tools, have proposed quantitative measures of individuality in order to facilitate their identification~\cite{krakauer2020information,biehlInformationBasedSpatiotemporal2016}, their algorithmic implementation remains difficult in practice. Second, among the very few works that proposed a deeper analysis of the robustness capabilities of the discovered patterns (based on the enumeration of all possible perturbations that a structure can receive from its immediate environment)~\cite{PMID:24494612,beer2020bittorio,cika2020resilient,agmon2015ontogeny}, findings suggest that glider-like structures typically remain quite fragile to external perturbations such as collision with other patterns~\cite{PMID:24494612}. 

In this work, we follow the enactivist framework and consider a class of continuous cellular automata called Lenia~\cite{chan2019lenia, chan2020lenia} as our artificial ``world''. We show that modern tools from machine learning can help scientists explore the vast space of continuous CA dynamics, enabling to address the problem of engineering robust second-order emergence. We propose a method based on curriculum learning, diversity search and gradient descent, enabling to efficiently shape the search process and to successfully navigate the chaotic outcome landscape of the high-dimensional Lenia system. 
In particular, we use a family of algorithmic processes called intrinsically-motivated goal exploration processes (IMGEP), an efficient form of diversity search algorithm~\cite{baranes2013active}. While mainly deployed in the fields of developmental robotics~\cite{Forestier2017IntrinsicallyMG} and developmental AI to enable robots explore and map vast sensorimotor spaces~\cite{colas2019curious, colas2020language}, recent works have shown how IMGEP can also form useful scientific discovery assistants for revealing the range of possible behaviors in unfamiliar systems such as chemical oil-droplet systems~\cite{grizouCuriousFormulationRobot2020}, physical non-equilibrium systems~\cite{falkCuriositydrivenSearchNovel2023} and models of continuous cellular automata systems as the one considered here~\cite{reinke2020intrinsically, etcheverry2020hierarchically}. At the difference of these previous works, we introduce two novel elements within the diversity search process: the use of gradient descent for local optimization and the use of stochastic perturbations within a curriculum of increasingly challenging and diverse target properties (hereafter called \textit{goals}).
With this method, we are able to find environmental rules leading to the emergence of patterns that self-constitute, self-maintain and move forward under various obstacle configurations, i.e. autopoietic entities displaying robust forms of sensorimotor agency. 

We then propose a battery of quantitative and qualitative tests, all formulated within the continuous CA paradigm, to further assess the robustness and generalization capabilities of the discovered self-organized patterns. Interestingly, the agents also show strong robustness to several out-of-distribution perturbations ranging from perturbing the agent structure in various ways not seen during training (including by a collision with another agent) to changing the scale of the agent. Furthermore, when tested in a multi-entity initialization and despite having been trained alone, not only the agents are able to preserve their individuality but they show forms of coordinated interactions (attractiveness and reproduction), which could be interpreted as a primitive form of intersubjective communication~\cite{PMID:24494612}. Those results illustrate the achievable generalization capabilities of artificial self-organizing agents, with respect to their mechanistic counterpart, opening interesting avenues for AI. At the same time, they provide interesting models about the way information might be processed by (brainless) biological agents to ensure robust maintenance of sensorimotor functions despite environmental and body perturbations~\cite{kitano2004biological}.

\section*{Study of sensorimotor agency in continuous CA models}
\begin{figure*}[t!]
\centering
\includegraphics[width=0.8\textwidth]{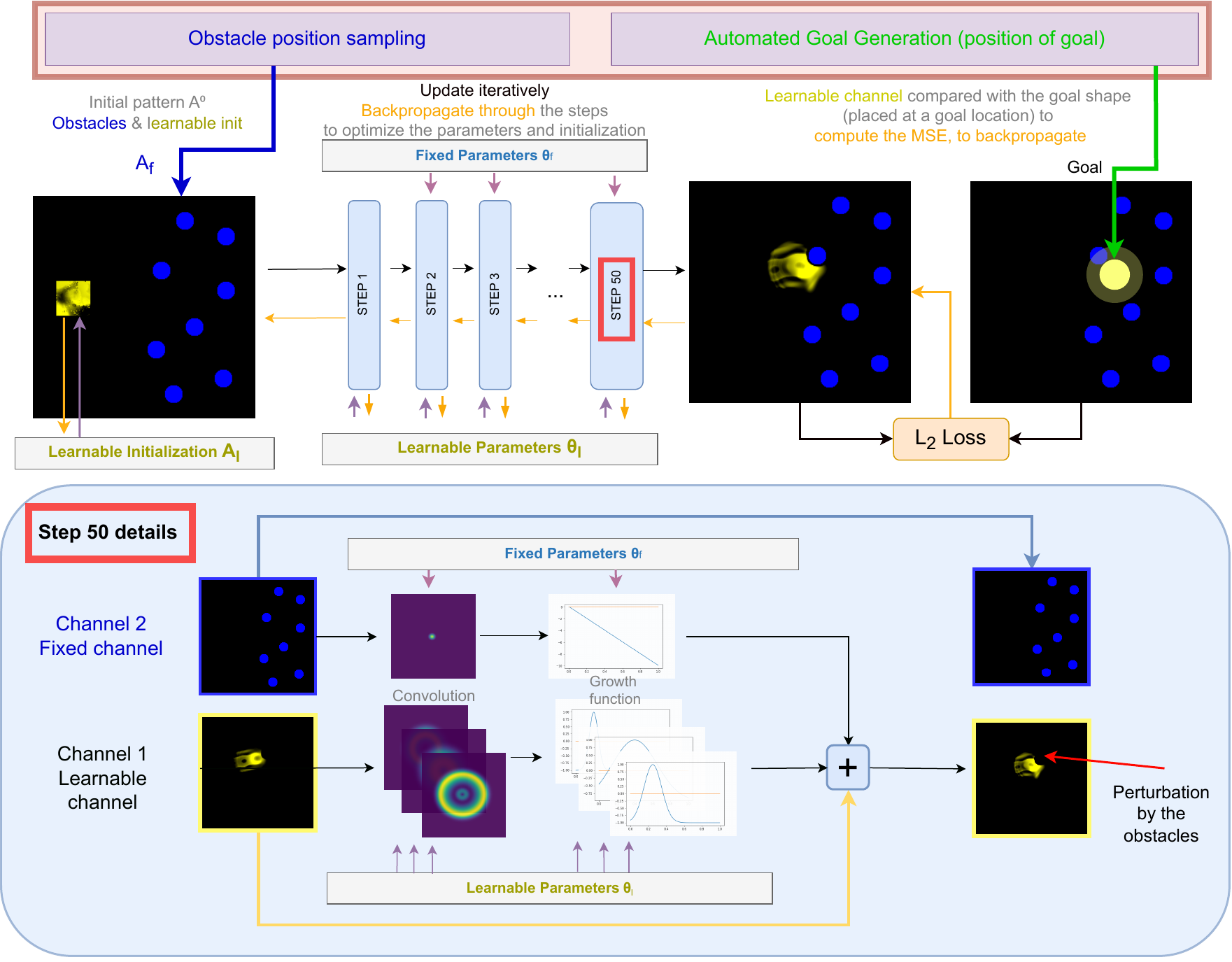}
\caption{\textbf{System overview}. (top) Illustration of one experimental rollout with automated (i) generation of target goal (green), (ii) generation of environmental obstacles (blue) and (iii) optimization of learnable parameters toward goal (backpropagation shown in orange). The initial state is iteratively updated by the parameterized rule, we then compute the goal conditionned loss from the last state of the rollout and propagate gradient across the steps to the learnable parameters and initialization. (bottom) Detailed view of a step in Lenia with obstacles. A convolution followed by a growth function is applied on each channel, resulting in a growth update which is added to the current state of the learnable channel. Both the convolution and the non-linear growth function in the learnable channel are parameterized (see appendix~\ref{fig:diff_lenia}).}
\label{fig:env_description}
\end{figure*}

Cellular automata (CA) are, in their classic form, a grid of “cells” $ A = \{ a_x \}$ that evolve through time $A^{t=1}\longrightarrow ... \longrightarrow A^{t=T}$ via the same local “physics-like” laws. More precisely, the cells sequentially update their state based on the states of their neighbours: $ a_x^{t+1}= f(\mathcal{N}(a_x^t))$, where $ x \in \mathcal{X}$ is the position of the cell on the grid, $a_x $ is the state of the cell, and $\mathcal{N}(a_x^t)$ is the neighbourhood of the cell (including itself). The dynamic of the CA is thus entirely defined by the initialization $ A^{t=1} $ (initial state of the cells in the grid) and the update rule  $f$ (how a cell updates based on its neighbours). But predicting the system long term behavior is a difficult challenge, even for simple rules, due to their potential chaotic dynamics \cite{WOLFRAM19841}.


In this work we use Lenia, a class of continuous CA which is a recently-proposed generalization of Conway's Game of Life~\cite{chan2019lenia, chan2020lenia}. Previous works in Lenia have shown that there exist local update rules $f$, that can lead to the self-organization of long-term stable complex patterns that display interesting life-like behaviors~\cite{chan2019lenia, chan2020lenia, etcheverry2020hierarchically}. Those include forms of individuality (spatially-localized organisation), locomotion (directional movement) and even basic behavioural capabilities (change of direction in response to interaction with other patterns in the grid). However, in previous work, self-maintenance of those behaviors in discovered spatially-localised patterns were typically quite fragile to external perturbations (for example collision with other agents \href{https://developmentalsystems.org/sensorimotor-lenia-companion/#orbium_collision}{Movie S3}), and properties of robustness and generalization were not specifically studied and tested: the possibility to self-organize robust self-maintaining ``agents'' was still an open question (and this applies to other CAs). Furthermore, these findings have so far relied on handmade exploration, which can be very hard and time-consuming as random rules rarely result in the emergence of localized patterns and even less moving ones (\href{https://developmentalsystems.org/sensorimotor-lenia-companion/#random_trials}{Movie S2}).     

In this work, we propose to use AI techniques to automate experimentation and the exploration of Lenia, with minimal human intervention. More particularly, the automated experimentation aims to find local update rules $f$ leading to the self-organization of stable (and if possible diverse) agents with sensorimotor capabilities. We also provide tests in order to assess the sensorimotor capabilities of the obtained patterns.

\subsection*{The Lenia environment}

Lenia is a class of continuous cellular automata where each CA instance is defined by a set of parameters $\theta$ that conditions the CA rule $f_{\theta}$. Once the parameters $\theta$ conditioning the update rule have been chosen, the system is a classical CA where the initial grid pattern $A^{t=1}$ is iteratively updated. In the multi channel version of Lenia \cite{chan2020lenia}, the system is composed of several communicating grids which we call channels. Intuitively, we can see channels as the domain of existence of a certain type of cell. Each type of cell has its own physics : it has its own way to interact with other cells of its type (intra-channel influence) and also its own way to interact with cells of other types (cross-channel influence).

In this work, we are interested in finding parameters ($\theta$,$A^{t=1}$) leading to the self organization of moving agents robust to external perturbations from the environment. For this aim, we need to introduce perturbations in the system in a controlled systematic way, both for testing the robustness and as criteria during the search. However, due to the dynamical nature of the system, controlled perturbations over several steps in the CA system are often hard to introduce. To help solve this issue, we propose to take advantage of the multi channel version of Lenia and separate the low level elements of the system in two types: the first ``fixed'' channel, which is hand-engineered, introduce elements that act as stable controlled obstacles (blue in Fig \ref{fig:env_description}); the second ``learnable'' channel, where parameters of the physic are learned, is where the agent has to emerge (yellow in Fig \ref{fig:env_description}). 
In practice, the environment parameters ($\theta$,$A^{t=1}$) are then separated in two. The first part, denoted ($\theta_f, A_f^{t=1} $) is a hand engineered part where $\theta_f$ gives the rule on how obstacles block matter from going in, while $A_f^{t=1}$ gives the obstacle placement and shape. Details on how we implement obstacles as part of the CA rule can be found in material and methods. The second part however, denoted ($\theta_l, A_l^{t=1} $), is free: the method presented below enables to learn these environment parameters so that ``agents'' with sensorimotor capabilities can self-organize.

What we are searching for is thus learnable parameters ($\theta_l, A_l^{t=1} $) that will induce a physic leading to the self-organization of agents that are able to move and survive in a grid where obstacles perturb their structure and therefore may break their integrity. Note that finding pattern with such capabilities is not trivial, for example moving patterns found by hand in \cite{chan2019lenia,chan2020lenia} (as the Lenia glider), which are stable without perturbations, often die from the collision with our engineered obstacles (\href{https://developmentalsystems.org/sensorimotor-lenia-companion/#orbium_obstacles}{Movie S4}). Note that in our system, if an agent is to emerge, the only way it can ``sense'' previously-introduced obstacles is from the perturbations that the obstacles induce on its structure. Compared to the physical world, the agent does not ``sense'' the obstacles by means of exchange of particles like photons or chemical molecules, as in vision or chemoreception, but more akin to direct touch as in haptic perception.

\subsection*{Intrinsically Motivated Goal Exploration Process (IMGEP)}

Formally, a set of parameters $(A_f,\theta_f,A_l,\theta_l)$ in Lenia maps to a certain sequence of states (trajectory $o$). This trajectory can then be mapped to a vector $R(o)$, through a defined characterization function $R$. This vector provides a behavioral description of the trajectory, and the image of $R$ represents the space of possible behaviors that can emerge in the system. As we will show below, randomly exploring the space of learnable parameters ($A_l,\theta_l$) is both costly in terms of experimentations, and inefficient for finding robust sensorimotor behaviour. 

Thus, we propose to leverage an AI technique called Intrinsically Motivated Goal Exploration Process (IMGEP) \cite{Forestier2017IntrinsicallyMG} to help exploring the space of behaviours. As this technique was originally developed to model curiosity-driven exploration in children \cite{gottlieb2018towards}, we call such a system a \textit{curious automated discovery assistant}. The IMGEP technique relies on \textit{goal-directed} search, which we leverage to drive the system toward the emergence of diverse target (sensorimotor) behaviors, called goals. More precisely, given a goal-sampling strategy $G$, IMGEP automatically samples target \textit{goals} $g \sim G$ which are points in the behavioral space. For each goal $g$, the objective is then to optimize toward parameters $(\theta_l,A_l)$ leading to a sequence of state which is mapped as closely as possible to this goal. To score the trajectory according to a goal, a loss function $\mathcal{L}(g,o)$ taking as input the trajectory and the goal is used.

 The behavioral descriptor $R$ we choose in this paper is the position of the center of mass at the last timestep of a simulation. The behavioral space then consists of all possible (x,y) coordinates in the grid. The objective for a given goal $g=(x,y)$ is thus to find parameters $(A_f,\theta_f) $ leading to the emergence of a spatially localized pattern attaining the goal position at the last timestep under several perturbation by obstacles. In this work, we choose to define the (goal-conditionned) loss as the mean squared error (MSE) between the state at the last timestep of the trajectory and a disk centered at the goal position. In addition to closeness to the goal position, the loss function we use incentivizes localization of the mass to prevent pattern explosion and collapse, which is a very common outcome of Lenia parameters. We then use gradient descent to optimize the learnable parameters ($\theta_l,A_l^{t=1}$) by backpropagating the loss through the steps and make progress toward the goal (Fig.\ref{fig:env_description}). 

Gradient descent optimization has already been successfully applied with cellular automata \cite{mordvintsev2020thread:} on learning CA parameters leading to the growth (and regrowth) of a target pattern~\cite{mordvintsev2020growing} or texture~\cite{niklasson2021self-organising}, or enabling cellular collectives to perceive their large scale structure \cite{randazzo2020self-classifying}, proving the effectiveness of such method (with some additional component for training for long term stability) in complex chaotic self-organizing dynamic. However, in this work, we consider moving agents which are a fragile type of pattern in Lenia as moving forward in such system means to grow new cells at the front while the ones at the back die. This equilibrium between growth and death is also challenged by the random perturbations we introduce in the system. This means that changes of parameters, because of the chaotic nature of the system, can easily break the equilibrium between growth and death of cells making the optimization harder. 

To help with this difficult optimization landscape we propose to introduce a \textit{curriculum} for making small improvements iteratively. Curriculum learning has already been applied for optimizing cellular automata rule with gradient descent as a solution for getting out of a trivial local optima in Variengien et al (2021) \cite{variengien2021SelforganizedControlUsing}. The curriculum also solves technical gradient flowing problem, detailed in appendix \ref{appendix:gradient}.

\begin{figure*}[t!]
\centering
\includegraphics[width=16cm]{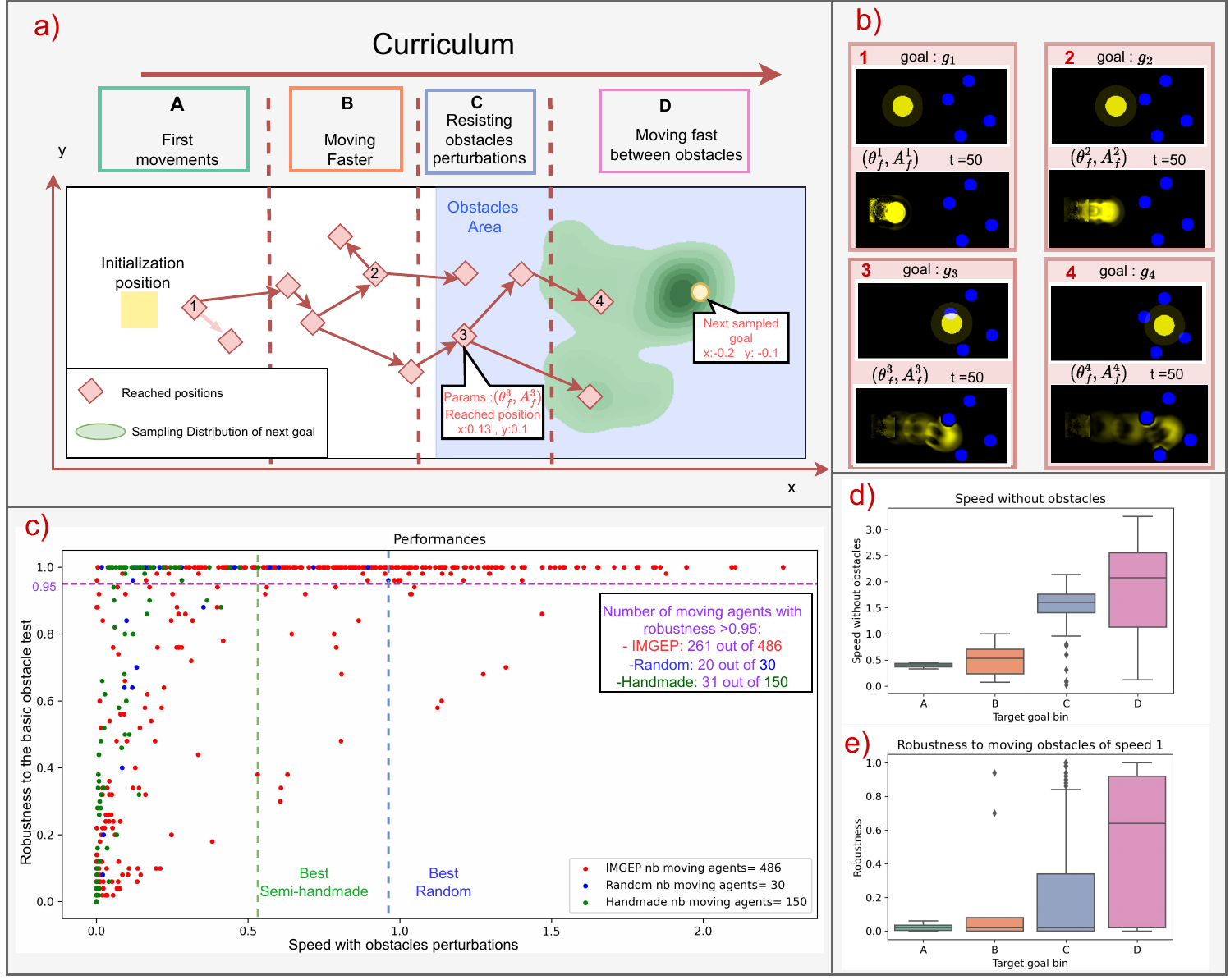}
\caption{\textbf{Curriculum and performances}. a) Schematic view of the curriculum. The curriculum iteratively sample goal positions (yellow disk), further in the grid, starting from very close to the initialization (A) to further away without obstacles (B) to further away in the obstacle area (C, then D). Arrow between reached positions (red square) represent that the parameters leading to a pattern attaining the tip of the arrow position was initialized before training by the parameters reaching the back of the arrow position. b) Examples of patterns obtained along the curriculum as well as their associated goal. We observe patterns going further and further in the same amount of steps (50 steps) and for the latter dealing with obstacles in their way. To display the trajectory of the agent in the learnable channel (yellow) we superposed the frames over all timesteps putting more transparency in earlier timesteps.  c) Performances in term of robustness to the basic obstacle test and speed with obstacle perturbations of the moving agent produced by: IMGEP (red), random parameters search with the same computation as our method, i.e. 117 000 parameters tried in total (blue) and handmade agents found in the original Lenia papers (green). d,e) Distribution of the Speed without obstacles perturbation (d) and robustness to moving obstacles (e) of moving agents obtained by the IMGEP along the curriculum. Details on these metrics can be found in Appendix.\ref{appendix:speed_measure},\ref{appendix:generalization_tests}. We observe that the curriculum is translated in an improvement in the 2 presented quantities.}
\label{fig:curriculum}
\end{figure*}

The intuitive idea behind our curriclum is to first learn rules leading to moving (spatially localized) agents which we train to go further and further (in the same amount of timesteps, hence faster) and at some point train them to go further while dealing with obstacles. To do so, the fixed environment $A_f^{t=1}$  we sample for training has a certain structure: the left half of the grid is free from obstacles while the right part contains obstacles that will be randomly placed at every rollout (blue in Fig.\ref{fig:curriculum}.a). The sampling strategy $G$ we chose in the IMGEP also participate to the curriculum as it is biased to randomly sample goals that are a little bit further than previously attained positions. More information on the sampling strategy can be found in appendix~\ref{appendix:goal_sampling}. Putting target goals in the obstacle area means that during training, the potentially emerging agents will have to go to a specific location while its structure is perturbed by obstacles randomly placed. The gradient descent optimization will incentivize recovery from perturbation and to keep moving despite being damaged. In addition, the fact that the obstacles are randomly placed should incentivize generalization to different perturbations.


To sum up, the IMGEP iteratively (and automatically) generates increasingly difficult goals, in increasingly difficult and diverse environments, for which we will try to find, and optimize using gradient descent, learnable parameters ($\theta_l,A_l^{t=1}$) that will lead to the self organization of agents achieving these goals. For each goal (position), the optimization steps are done under several obstacle configurations $\{A_f\}$ in order to learn to resist to different perturbations. After each optimization, we then test the final obtained parameters on several obstacles configurations $\{A_f\}$, that are sampled the same way as in the training steps, to assess the reached position. We store this (parameters, reached position) couple in history $\mathcal{H}$ in order to be able to use it as a starting point for subsequent goals. A more detailed description of the method can be found in material and methods.


\subsection*{Evaluation of the discovered patterns}
Whereas the notion of \textit{agency} is closely tied to the ability of an organism to maintain its own organization despite encountering novel circumstances, the robustness of current artificial autopoietic systems is lagging far behind the robustness of their biological counterparts. 
We believe that this limitation, together with the difficulty of engineering such autopoietic systems, is a major reason why we have not assisted yet to a wider adoption of the enactivist framework by the AI community.
The IMGEP search, which is precisely intended to facilitate the search of such autopoietic systems, should provide us with a database of parameters $\{(A_f, \theta_f)\} \in \mathcal{H}$ that (when successful) lead to the self-organization of patterns that are robust (at least) to the different obstacle configurations seen during training. 

To go further and characterize \textit{agency} and the degree of \textit{robustness} of the discovered parameters/patterns, we propose an empirical evaluation procedure in two stages. 
First an ``empirical agency filter'' is used on the database of discoveries to discard parameters that do not lead to the self-organization of what we call ``agents'' in Lenia. More precisely, our filter implements several classifiers, inspired from ones proposed by Reinke et al.~\cite{reinke2020intrinsically}, to detect whether the emergent matter does not disintegrate (vanishes or explodes), forms a coherent entity (single soliton), and does so during a long-enough time window (longer than training). In addition to the agency filter, we also introduce a moving filter which tells if an agent is moving (travels a minimum distance) or not (examples of discovered ``agents'' that are considered not moving are shown in \href{https://developmentalsystems.org/sensorimotor-lenia-companion/#non_moving}{Movie S19}). Then, to assess the capabilities of selected agents to withstand perturbation by obstacles we perform a basic obstacle test: testing them on obstacle configurations similar to the ones seen during training; and various generalization tests: running them through a battery of tests with several \textit{out-of-distribution} perturbations that were not seen during training. In particular, we test the discovered sensorimotor agents to harder obstacle configurations, stochastic cell updates, changes of initialisation and changes of scale that were not experienced during training. For each test, given a distribution of perturbations, we measure \textit{robustness} as the average performance over sampled perturbations, where performance is a binary success metric that determines whether the agent ``survived'' the perturbation or not. As for ``survival'' metric, we simply apply our agency filter to detect whether the (perturbed) emergent entity is able to self-maintain despite the introduced perturbations (i.e. is still an agent at the end of the test). Note that this metric closely follows the definition of \textit{cognitive domain} of an autopoietic system, which was introduced by Maturana and Varela~\cite{maturana1980autopoiesis} and later defined by R. Beer as the percentage of \textit{non-destructive}\footnote{a perturbation is said to be \textit{destructive} if it fundamentally disrupts the entity’s organization leading to its disintegration~\cite{beer2020bittorio}} perturbations, out of all possible perturbations, that the autopoietic system can tolerate~\cite{beer2020bittorio}. Because measuring the cognitive domain as such would require an exhaustive enumeration of all possible perturbations and all possible valid states that the entity can take, which is not tractable in the Lenia environment, we instead rely on a proxy metric and on a set of chosen empirical tests. Finally, in addition to robustness,  we also measure the performance of agents in term of \textit{speed} with and without obstacles, especially as speed can be a measure of performance of motor capabilities (for example for biological agent to flee predators or chase preys) and as speed with obstacles is an interesting measure on how well the agent deals with obstacles. We refer the reader to Material and Methods and to appendix \ref{appendix:tests} for more details on our evaluation procedure.

In addition, we provide the code\footnote{Source code for reproducing the results can be found at \url{https://github.com/flowersteam/sensorimotor-lenia-search}.} enabling to reproduce all results, as well as an interactive web-demo\footnote{Interactive web demo and additional videos can be found at \url{http://developmentalsystems.org/sensorimotor-lenia-companion/}} where one can replay the discovered agents and test them to all sorts of freely-drawn perturbations including  custom obstacle shapes, addition and/or removal of mass, interactions with other agents in the grid and control of environmental cues (attractive elements) in the Lenia grid. 

We argue that those quantititative and qualitative tests, which were all implemented within the continuous CA paradigm, can serve as a good baseline to evaluate the generalization capabilities (and hence the degree of agency) of autopoietic systems in enactivist research, akin to commonly deployed benchmarks in AI for evaluating mechanistic forms of agency~\cite{team2021open}.

\section*{Results}

In this section, we analyze the discoveries made by the proposed approach (IMGEP) and compare it with two other exploration baselines: a \textit{random search}, where parameters are sampled uniformly in the parameter space (same ranges than for the IMGEP, given in appendix \ref{appendix:lenia_params}); and a \textit{handmade search}, where we collected the discoveries, made by semi-automatic search and expert selection, presented in the original Lenia papers \cite{chan2019lenia,chan2020lenia}. 
Each IMGEP experiment outputs 160 parameters but performs in average 11700 Lenia rollouts, due to stochasticity in the method (see Materials and Methods). For IMGEP and random search, 10 independent repetitions are performed (where random search is given the same experimental budget of 11700 rollouts per seed). Note that the comparison with handmade search, while interesting, is challenging in practice as it is the result of tedious search for which the total experimental budget is unknown, and which was conducted over some Lenia hyper-parameters that are not all included in the automated search (e.g. various number of channels or kernels). Moreover we use a slightly different parameterization of the rule to allow for differentiability (details in appendix \ref{appendix:lenia_diff}). 

For the three baselines (IMGEP, random search and handmade search), we filter the obtained parameters to select only the moving agents (passing the agency and moving test) and measure their speed and robustness to the basic obstacle test and generalizations tests, as described in previous section.

\subsection*{Individuality, locomotion and sensorimotor capabilities}
\label{subsec:individuality-locomotion-sensorimotor}

As illustrated in Fig.\ref{fig:curriculum}, the IMGEP search enables to evolve agents along a curriculum which progressively lead to the emergence of individuality, locomotion and sensorimotor capabilities. At first, the IMGEP samples goals (i.e. target positions) that are not too far from initialization (area A in Fig.\ref{fig:curriculum}.a) and enabling to find rules leading to the self organization of spatially localized patterns which starts to move a little bit from initialization (as shown in Fig.\ref{fig:curriculum}.b.1). Then, from these newly learned rules the IMGEP samples further goals (area B in Fig.\ref{fig:curriculum}.a) which lead to spatially localized patterns that move further in the grid in the same amount of time (Fig.\ref{fig:curriculum}.b.2). At this point, some obtained parameters already lead to the self-organization of \textit{moving agents} i.e. passing our empirical agency test and moving tests (long-term stable solitons capable of moving while self-maintaining). Moving agents patterns are in fact already not trivial to find through random search in the parameter space as only 30 moving agents were found through the 10 seeds of random search out of a total of 117 000 trials of parameters. The speed of the obtained moving agent at this point is still limited as can be seen in Fig \ref{fig:curriculum}.d.

The IMGEP pursues the curriculum, taking advantage on the previous learned parameters that already result in moving agents, now sampling target goals that are even further away from the initial position, in the obstacle area C,D in \ref{fig:curriculum}.a, leading to moving agents entering the obstacle area (as shown in Fig.\ref{fig:curriculum}.b.3,4). As expected, the parameters resulting from those goals have a higher robustness to obstacles as can be seen in Fig.\ref{fig:curriculum}.e. We refer to appendix \ref{appendix:imgep_no_obstacles} for extra experiment with an ablation of the obstacle area during optimization showing that the increase of robustness is due to the presence of obstacles in the optimization and not only to the the distance of the target goal position to the initialization.


As expected, we observe that agents trained with further goals move in average at faster speeds in environment without obstacles (Fig.\ref{fig:curriculum}.d)

At the end of the curriculum loop, the obtained rules often lead to the self-organization of moving agents that are able to navigate fast in an area with obstacles while still maintaining their integrity (Fig.\ref{fig:curriculum}.b.4, \href{https://developmentalsystems.org/sensorimotor-lenia-companion/#sensorimotor_agents}{Movie S1}). The emerging agents are capable of changing direction and recover in response to perturbations induced by the obstacles, i.e. have sensorimotor capabilities, and this only through the global coordination of those identical low level parts and in particular without having any central unit computing decision. 

In total, 9 out of the 10 seeds led to at least one sensorimotor agent, which we define in this paper as moving agent with a measured robustness >0.95 in our basic obstacle test. Note however that the performance in term of speed with obstacles varies from one seed to another (see Appendix Table \ref{appendix:seed_var_tab}).


Over the 10 seeds a great part of the obtained emerging moving agents are sensorimotor agents. In fact, over 10 seeds, 486 of the 1600 parameters (10 seeds x160 parameters) led to moving agent according to our empirical agency and moving filter, from which 261 have a robustness to obstacles >0.95.


 As a comparison, out of the 117 000 parameters generated by the 10 seeds of random search, only 30 led to moving agents from which 20 have a robustness to obstacles >0.95. 
Our method surpasses random search in term of speed with obstacles and robustness of the obtained agents, as well as the total number of long term stable moving agents obtained as can be seen in Fig.\ref{fig:curriculum}.c (486 for IMGEP and 30 for random search in total over 10 seeds and with the same Lenia rollout budget). In fact random search is able to find some agents ($\sim 1\%$ of all its discoveries) but most of them are static compared to IMGEP whose directed search fosters the emergence of moving agents (\href{https://developmentalsystems.org/sensorimotor-lenia-companion/#overview_agents}{Movie S5}).

Our method also results in agents with better robustness and speed than the ones found in the original Lenia papers \cite{chan2019lenia,chan2020lenia} (Fig.\ref{fig:curriculum}.c). 

Ablation studies of the method can be found in the appendix~\ref{appendix:ablation}, showing how curriculum, diversity search and gradient descent are key ingredients in the method and are an efficient direction to search for sensorimotor behavior in self-organizing systems. We also provide the sequence of reached positions of a seed in appendix \ref{appendix:Phylogeny}, displaying the curriculum and showing how diversity search can help find potential stepping stones.

\subsection*{Generalization}
\begin{figure*}[t!]
\centering
\includegraphics[width=0.85\textwidth]{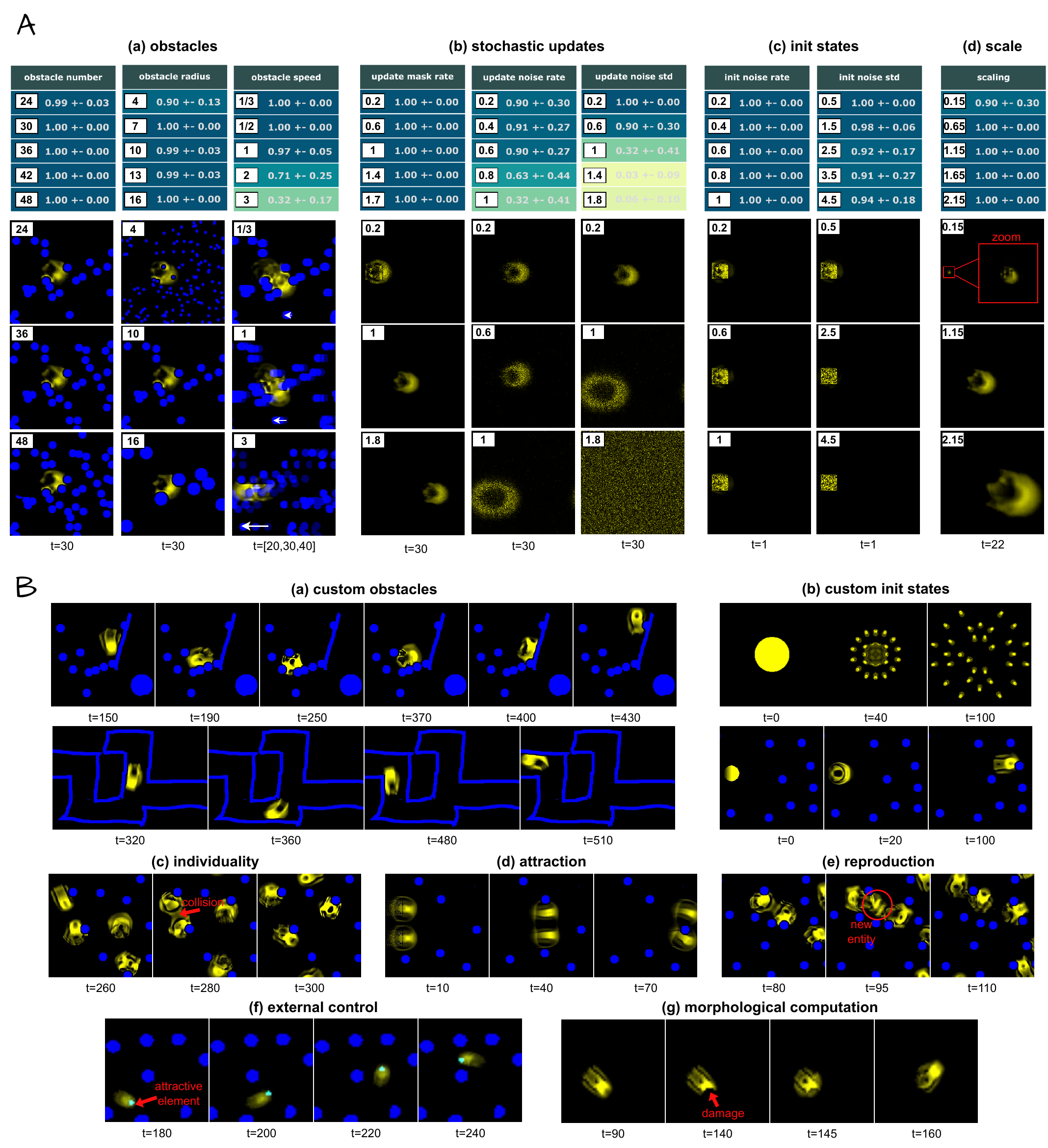}
\caption{\textbf{Generalization of the discovered sensorimotor agents}. (A) We conduct a battery of quantitative tests which we organize in 9 families of parameterized perturbations that test for various (a) obstacle number, size and speed, (b) rate of cell updates, as well as rate and magnitude of noise added to the updates, but also (c) rate and magnitude of noise added to the initial state and (d) scaling factors. For each family, we test for $5$ different parameter values, i.e. perturbation strength, resulting in a total of $9\times 5 = 45$ tests. For each test, the performance of an agent is computed as the average  score of survival over 10 random seeds. A score of 1 (dark blue) means that the agent survived all 10 tests whereas a score of 0 (light yellow) means that the agent survived none of the tests. The table reports the mean and standard-deviation performances, over the 10 best agents discovered by our goal-directed curriculum, for all of the 45 tests (one table cell per test), where ``best'' is determined by the speed/robustness criteria introduced in Figure \ref{fig:curriculum}-c. Below each column, we show snapshots of system rollout at test time given the newly introduced perturbations. The shown snapshots are all taken from rollouts of the ``best'' agents, and from the first seed (out of the 10 tested random seeds).  Timesteps are specified under the images, for instance snapshots of the perturbations applied on the initial state are shown at t=1.
(B) We also conduct a battery of qualitative tests, where we tested the (best) discovered agents to all sorts of difficult perturbations including (a) freely-drawn obstacles such as walls, mazes or dead-ends (b) freely-drawn initial states such as very big disks (resulting in the emergence of multiple entities) or small disks with gradient asymmetry, (c-d-e) introduction of other agents in the grid (resulting in the emergence of inter-agent interactions such as individuality maintenance, attraction and reproduction), (f) the introduction of novel low-level elements that have an ``attractive'' effect on the agents (allowing external user to guide the agent trajectory in the grid); and (g) custom mass removal (pixel erasing). Details of the resulting observed behaviors are provided in the text, with videos available on the companion website \url{https://developmentalsystems.org/sensorimotor-lenia-companion/}.}
\label{fig:generalization}
\end{figure*}

Biological organisms are able to maintain phenotypic stability in the face of diverse environmental perturbations arising from external stresses, intracellular noise, and even quite drastic changes during morphogenesis such as perturbations to the embryo structure~\cite{vandenberg2012normalized} or to the substrate cellular size~\cite{fankhauser1945maintenance}. It has long been recognized that robustness is an inherent property of all biological systems that has been strongly favored by evolution~\cite{stelling2004robustness}. 
In this section, we are interested to see if similar robustness capabilities can be achieved by the artificial self-organizing agents that have been discovered by our artificial evolution workflow (Figure \ref{fig:generalization}). To do so we evaluate the generalization capabilities, over the proposed battery of tests, of the 10 best agents discovered by the IMGEP, random and handmade search variants, as well as on the agents that have a speed within obstacles greater than one (91, all discovered by IMGEP). ``Best'' here is computed according to the speed-robustness criteria presented in Figure \ref{fig:curriculum}-c, i.e. the fastest with obstacle that also have a robustness in the basic obstacle test>0.95. The performances are fully reported and compared in appendix tab.\ref{appendix:generalization_table}. As we will see, the discovered agents showcase quite impressive generalization capabilities at the organic, sensorimotor and inter-subjective levels~\cite{di2019process}. We group the observed generalization capabilities into six categories: harder obstacle configurations (external stresses), stochastic cell updates (per-cell noise), changes of initialisation (``embryo'' variation), changes of scale (compute capacity variation), interactions with other agents in the grid (inter-agents regulation) as well as with human-controlled environmental cues (observer-agent regulation). 

\subsubsection*{Harder obstacles} We first tested the agents generalization capabilities to a larger and more challenging set of obstacle configurations. The test set includes controlled configuration with varying number, size and speed of obstacles (Figure \ref{fig:generalization}-A-a), as well as human-drawn obstacles such as vertical walls and dead ends (Figure \ref{fig:generalization}-B-a).
Interestingly, whereas some well-placed perturbations can lead to death or explosion, the discovered agents show strong robustness and generalization to most of the test set configurations. They showed quasi-perfect survival to grids with up to 48 obstacles, to grids with small (but dense) or big (but sparser) obstacles, and to obstacles with moderate speed. High-speed obstacles however, seem to challenge agent's survival (Figure \ref{fig:generalization}-A-a), even though the IMGEP-discovered agents are still much more robust to moving obstacles than the ones discovered by random and handmade search (appendix Table \ref{appendix:generalization_table} and \href{https://developmentalsystems.org/sensorimotor-lenia-companion/#moveobs_overview}{Movie S6}). Those results suggest that, by training for fast-moving and obstacle-resisting behaviors, our goal-directed curriculum favored the self-organization of agents that are able to \textit{quickly} recover from perturbations induced by the environment, even ones not seen during training. 
For instance, qualitative tests also showed that the discovered agents are able to successfully navigate forward while coming across tightly-packed obstacles, walls of various inclinations, corners, dead ends and even bullet-like types of obstacles (\href{https://developmentalsystems.org/sensorimotor-lenia-companion/#bullets}{Movie S10}).

\subsubsection*{Stochastic updates} We then tested the agents generalization capabilities to asynchronous and noisy cell updates. As proposed in Mordvinstev et al.~\cite{mordvintsev2020growing}, relaxing the traditional assumption of synchronous update in cellular automata (which assumes a global clock) is closer to what you would expect from a self-organized system, and can be done by applying a random update mask on each cell (parameterized by the update mask rate in Figure \ref{fig:generalization}-A-b). Despite the update mask enforcing asynchronous and less (or more) frequent cell updates at test time, the discovered parameters still give rise to self-organized agents that perfectly self-maintain (survival scores of one) and that showcase very similar morphology and behavior as the agents with synchronous updates (\href{https://developmentalsystems.org/sensorimotor-lenia-companion/#asynchro}{Movie S14}). The agents are slowed (or fasten) a little bit but this is what we can expect as each cell is updated in average only a fraction of the time (or several times per timestep). We also relax the assumption of exact update by adding random noise, of various amount and magnitude, to the cell states during the system rollout. Here, we observe that the agents can resist quite consequent quantities of noise but passed a certain level, as expected, the collective looses its integrity and disintegrates (Figure \ref{fig:generalization}-A-b).

\subsubsection*{Changes of initialization} 
While the initialization pattern has been learned with a lot of degree of freedom (pattern in $[0,1]^{40\times40}$), we can look if similar patterns (phenotypes) can self-organize from other (maybe simpler) initialization patterns. This capacity to converge to the desired anatomy in spite of a different initialization (``embryo''), is something that can be found in biological organisms~\cite{vandenberg2012normalized}, and that we can expect in our system as well. Indeed, as shown in Figure \ref{fig:generalization}-A-c, we can see a quasi-perfect robustness to noise-altered initial states, and this even for quite high amounts of noise (except for few configurations that lead to death). These results suggest that the final phenotype forms a strong attractor towards which the different initial mass pattern tend to converge under the learned CA rule. The learned CA rules are hence prone to encode, grow and maintain a specific target morphology (and its associated functionality), which is consistent with the agent ability to recover from obstacle-induced perturbed morphology. As illustrated in Figure \ref{fig:generalization}-B-b, we also tested for handmade initial patterns such as bigger disks and same-size asymmetrical disks (for example with gradient activation). Interestingly the large disk initialization led to multiple entities forming and separating from each other. The same-size disk, which is much simpler than the trained initial states (but preserves some form of asymmetry) also converged toward the same morphology. However the robustness to initialization is not perfect as many initializations, such as disk of smaller size and/or without asymmetry, easily lead to death (\href{https://developmentalsystems.org/sensorimotor-lenia-companion/#init_death}{Movie  S18}).

\subsubsection*{Changes of scale} Similarly, while the initialization and update parameters have been learned at a certain spatial resolution during training resulting in agents of a certain size (in term of number of cells), we can artificially change the scale at test time by approximate resizing of parameters (see Appendix section \ref{appendix:generalization_tests}). As shown in Figure \ref{fig:generalization}-A-d, we tested for different down-scaling (and up-scaling) factors that surprisingly resulted for most of them in fully functional agents with the overall same structure but smaller (or larger) size in terms of number of cells. For agents which are down-scaled, and hence have much less pixels/cells to do the computation, it is particularly surprising that they are still able to sense and react to their environment and still show relatively-advanced levels of robustness (\href{https://developmentalsystems.org/sensorimotor-lenia-companion/#scaling}{Movie S15}). This scale reduction has a limit (a scaling of 0.15 already leads to some death) but we can go quite far down and still obtain functional phenotypes. For the bigger agents, which therefore have more space to compute (but also more cells to organize), we observe similar results where agents still self-organize to functional phenotype. Once again, this resonates with findings in biology suggesting that organisms are able to accommodate cell-size differences by adjusting cell number in order to maintain roughly constant body size and structure~\cite{fankhauser1945maintenance}. 

\subsubsection*{Interactions} We were then interested to test how the discovered agents would react when interacting with other agents in the grid. Given the set of parameters ($A_l,\theta_l)$, we can trigger the forming of several macro-entities at test time by replicating the initialization square pattern ($A_l \in [0,1]^{40 \times 40}$) at different locations within a larger grid ($A^{t=1} \in [0,1]^{512 \times 256}$) and letting the system unroll. Doing so leads to the development of several entities of the same ``specie'' (governed by the same update rule/physic $\theta_l$). As illustrated in Figure~\ref{fig:generalization}-B, we did that for several of the discovered sensorimotor agents, and qualitatively observed several interesting emergent interactions.

The first thing that we observed is that, several of the discovered agents show strong \textit{individuality} preservation (\href{https://developmentalsystems.org/sensorimotor-lenia-companion/#individuality}{Movie S11}). The fact that the individual agents do not merge nor enter in destructive interactions despite being all made from identical cells is an intriguing example of how the boundary of a ``self''~\cite{levin2019computational} can emerge and maintain in self-organizing systems. In particular results suggest that, in the Lenia system, individuality can be obtained as a byproduct of training an agent alone. Our intuition is that by trying to prevent too much growth during training, it learned to prevent any living cell that would make it “too big”, including living cells from other entities here. 

A second type of interaction that can be observed with certain parameters/environments is \textit{attraction}. As illustrated in \href{https://developmentalsystems.org/sensorimotor-lenia-companion/#attraction}{movie S13}, two agents placed in the same grid can show attraction when coming close enough from one another, leading them to stay together and move in the same direction. Interestingly, when they encounter an obstacle, they are able to separate briefly and then to reassemble together. Similarly, even when they stay together, we can still qualitatively observe two distinct entities that are interacting with one another while maintaining their overall shape and integrity. This type of behavior has been studied in the game of life under the concept of \textit{consensual domain}~\cite{PMID:24494612}. 

A third type interaction that has been observed in some of the discovered agents is a form of \textit{reproduction} where collision between two agents give rise to the birth of a third entity (\href{https://developmentalsystems.org/sensorimotor-lenia-companion/#reproduction}{Movie S12}). This kind of interaction seems to happen when one of the two colliding entities is in a certain “mode”, like when it just hit a wall. Our intuition is that when it hits a wall, the self-organizing agent produces a growth response in order to recover. During this growth response if there is extra additional mass coming from another entity then the self-organizing agent might split off from the created mass while the separated mass, from robust self-organization (see ``Changes of initialization'' above), grows into a complete individual.

\subsubsection*{External control} A central challenge in synthetic biology, when faced with unconventional forms of agency such as collective of cells, is to find new ways to communicate with the cells to induce desired behaviors at the collective level without having to physically ``rewire'' the structure of the agent (e.g. via genome editing) but rather by introducing externally-controlled cues in the environment~\cite{levin2022technological}. Here, we are interested to see whether we can induce (novel) target behaviors in the discovered agents without having to modify the learned parameters $\theta_l$. In particular, we investigate whether the agents can show \textit{attraction} to some novel elements in their environment (like in nature organisms being attracted to certain chemicals, lights or temperatures) and if we could use those elements to guide the macro-entity. To do so, we introduce a new type of ``attractive'' low-level elements within the Lenia CA paradigm. More precisely, given the set of learned parameters $\theta_l$, we introduce a novel local rule with parameters $\theta_a$ that determine the physical influence of the attractive elements onto the agent cells. To find parameters $\theta_a$ triggering the desired attraction effect at the agent behavioral level, a simple random search with automatic pre-filtering and final human assessment was performed (see appendix \ref{appendix:attraction} for details on the procedure). \href{https://developmentalsystems.org/sensorimotor-lenia-companion/#attraction_external}{Movie S17} is an example of obtained behavior where we can clearly see that the sensorimotor agent is getting attracted to the newly-introduced environmental element (disk of cyan particles) which allows the external user to ``control'' the agent trajectory by moving the disk in the grid. Interestingly, in spite of this novel behavior, agents are capable to maintain their normal sensorimotor capabilities showing robustness to collision with obstacles and other agents in the grid. Besides, once the attractive element is removed the agents return to their normal behavior. However adding extra rules also fragilize equilibrium that existed in the agent rules as it creates perturbations that the agent has not been trained to withstand, leading sometimes to death or explosion (or to other behaviors such as reproduction due to extra boost of growth). Once again parallels can be drawn with findings in biological organisms, for instance \cite{li2016light} show that controlled UV light beam can be used to externally guide the trajectory of micro-swimmers to perform on-demand drug discovery. While we only tested for attraction-type of generalization behaviors, we believe that more sophisticated types of environmental guidance could be induced, though probably necessitating more advanced search methods.

\subsubsection*{Morphological computation} This section has provided several empirical evidences of how adaptive high-level functionality can emerge from a collective of low-level, decentralized elements. In order to withstand the tested perturbations, the cellular collective first needed to ``sense'' the induced perturbations through a deformation of the macro structure. After this deformation it had to ``communicate'' the information and make a collective “decision” on where to grow next. Then it had to move and regrow its shape, altogether giving rise to the observed robustness of the macro structure. In order to better visualize the physical manifestation of decision-making within the cellular collective, we manually suppressed a part of the agent (Figure \ref{fig:generalization}-B-g, \href{https://developmentalsystems.org/sensorimotor-lenia-companion/#damage_hand}{Movie S16}). We can clearly observe that perturbation of the macro-structure is what leads to the direct change of direction. Those results support the fact that computation of the decision is made at the morphological level hence that morphology, decision-making and motricity are highly entangled phenomena~\cite{pfeifer2006body}. 

\section*{Discussion}
In closing this paper, let us reiterate that what is interesting in such a system is that the computation of decision is done at the macro (group) level, showing how a group of simple identical entities can make “decision” and “sense” at the macro scale through local interactions only, and without a clear pre-existing notion of body/sensor/actuator. Seeing the discovered agents, it’s even hard to believe that they are in fact made of tiny parts all behaving under the same rules. While some basic behavioural capabilities (spatially localized and moving entities) had already been found in Lenia with random search and basic evolutionary algorithms, this work makes a step forward showing how Lenia’s low-level rules can self-organize robust sensorimotor agents with strong adaptivity and generalization to out-of-distribution perturbations. 

Moreover, this work provides a more systematic method based on gradient descent, diversity search and curriculum-driven exploration to easily learn the update rule and initialization state, from scratch in high dimensional parameters space, leading to the systematic emergence of different robust agents with sensorimotor capabilities. We believe that the set of tools presented here can be useful in general to discover parameters that lead to complex self-organized behaviors. 

Yet, several of the analyses we make in this work are empirical estimations or subjective. Future work shall consider how more formal definition(s) of agency and sensorimotor capabilities could be applied to the high-dimensional systems studied here\cite{krakauer2020information,biehlInformationBasedSpatiotemporal2016}.

Also, engineering subparts of the environmental dynamics with functional constraints (through predefined channels and kernels) has been crucial in this work to shape the search process \cite{DBLP:journals/corr/abs-1710-03748} towards the emergence of sensorimotor capabilities, as well as used as a tool to analyze more easily these emergent sensorimotor capabilities. An interesting direction for future work is to add even more constraints in the environment such as the need for food/energy to survive, the principle of mass conservation, or even the need to develop some kind of memory to anticipate future perturbations. We believe that richer environmental constraints and opportunities might be a great leap forward in the search for more advanced agent behaviors. For example, behaviors like competition between individuals/species for food, foraging or even basic forms of learning might emerge. From this competition and new constraints, interesting strategies could emerge as a form of autocurricula, as in \cite{baker2020emergent,DBLP:journals/corr/abs-1710-03748}. Promising steps in this direction have been made in \cite{plantec2023flowlenia} which introduces Flow Lenia: a mass conservative version of Lenia. In particular, Flow Lenia also allows several species of agents to coexist in the same grid, leading to competition for matter between the species.

In fact, beyond individual capabilities, we could even wonder under what conditions one could observe the emergence of an open-ended evolutionary process \cite{stanley2017open} directly in the environment, without any outer algorithm, resulting in the emergence of agents with increasingly complex behaviors. Like building the physical rules of an “Universe” and letting agency and evolution emerge from the interactions between parts. To achieve this, we might need to use an optimization process similar to the one presented in this article to evolve all the environmental rules instead of pre-specifying some of them by hand. Indeed, while the engineering of specific environmental rules facilitates the understanding/studying of the results, having more systematic ways to generate them could take us closer to the fundamental scientific quest of designing open-ended artificial systems with forms of functional life and agency “as it could be”. Some preliminary studies are underway \cite{chan2023largescale}.

Beyond those fundamental scientific questions, future work might also consider broader applications of this work for biology and AI. In biology, inferring low-level rules to control complex system-level behaviors is a key problem in regenerative medicine and synthetic bioengineering \cite{pezzulo2015re,pezzulo2016top}. In this regard, cellular automata offer an interesting framework to model, understand and control the emergence of growth, form and function in self-organizing systems. However, they remain abstract models: entities in the CA exist on a predefined grid topology whereas physical entities have continuous position and speed ; states in the CA are well-defined whereas it is not clear where and how information is processed in living organisms; rules in the CA operate at a predetermined scale whereas real-world processes operate at nested and interconnected scales.
In AI, with the recent rise of web-deployed machine-learning models including large language models~\cite{nakano2021webgpt, schick2023toolformer}, we are also faced with an increasing blurring of boundaries between the AI and the rest of the ``environment'' (human end-users and the web itself). It is hence central to understand how to measure emergent agency and cognition in those AI systems, as well as how to interact with them despite the extremely large input and behavioral spaces involved. In this regard we believe that environments like the one considered in this work can be useful to better inform the debate in much bigger models, as they are rich enough to support emergent agential behaviors while simple enough to study those questions explicitly. Far from trivial, transferring insights from  the considered artificial systems to real biological systems or to very large AI systems is an exciting area of research with a potential broad range of medical and societal applications \cite{kriegman2020scalable,ebrahimkhani2021synthetic}.


\matmethods{

\subsection*{System}\label{matmethods:system} An update in Lenia is given by the different rules composing the function $f_{\theta}$, each rule is composed of a convolution kernel (which will sense the surrounding of the cell) and a growth function (a function which will convert this sensing, a scalar, into an update of the mass, another scalar). The update of the cells are then given by a weighted sum of the update given by each rule. At each step, the calculation of the update is done identically on every cell of the grid (every cell apply the same convolution filter and growth function). This update is then added on the associated cell and the result is clipped between 0 and 1. See figure \ref{fig:env_description} for an illustration of the update. The Lenia system used in this work is slightly different from the one in the original paper \cite{chan2019lenia,chan2020lenia}. We changed the parameterization in order to allow more gradient to flow through the steps (more details in appendix \ref{appendix:lenia_diff}). We also choose to use 10 rules, from the learnable channel to itself. We refer to appendix \ref{appendix:lenia_params} for a detail on the parameter of the systems and their role. In total the 10 rules are controlled by 132 parameters.

\subsection*{Modeling of Environmental Constraints}\label{matmethods:obstacles}  The parameter $\theta_f$ gives the update rule associated with obstacles. This rule senses in the obstacle channel and update in the learnable channel. This means that the convolution will be calculated upon the obstacle channel and the growth obtained through the growth function will be added to the learnable channel. In practice, for $\theta_f$, we use a rule with a convolutional kernel of small size, so that obstacles have effects only locally and a growth function which has a huge negative decrease of mass in the learnable channel to prevent any matter from going where obstacles are present. More information in appendix \ref{appendix:lenia_obstacles}.

\subsection*{IMGEP} \label{matmethods:IMGEP} 

Our proposed method, based on the IMGEP framework \cite{Forestier2017IntrinsicallyMG}, and fully described in appendix \ref{appendix:imgep_details}, starts by initializing the history with 40 random parameters and their associated reached position (position of the center of mass at last timestep) computed over 20 rollouts with random obstacle configurations. The method then begins a loop where each step is composed of 1) the sampling of a goal (x,y position in the grid) , then 2) a selection from the history of the parameters reaching the closest goal which will be used to initialize the parameters, 3) an optimization of those paramaters towards the goal under several obstacle configurations, 4) a test of those parameters over 20  obstacle configurations to compute the final reached position after optimization, and adding the couple (parameters, reached position) to the history to reuse it in next steps. Pseudo code \ref{algo:ad_framework} and figure \ref{appendix_fig:IMGEP} illustrating the IMGEP algorithm can be found in appendix. Details of each step of the method: 1, 2, 3, 4 can be respectively found in appendix \ref{appendix:goal_sampling}, \ref{appendix:history_reuse_filter}, \ref{appendix:gradient}, \ref{appendix:param_eval}.


In this work, the loop defined above is composed of 120 outer steps where 1 out of 5 outer steps performs 125 steps of gradient descent while the rest performs random mutation on the initialized parameters and 15 steps of gradient descent (details on mutations in appendix \ref{appendix:mutation_details}). At every gradient descent step (Fig.\ref{fig:env_description}), we run a Lenia rollout with the current parameters ($\theta_l,A_l$) and random obstacle placement ($A_f$) for 50 timesteps and apply a mean square error loss between the last state of the learnable channel (at last timestep) and a disk centered at the position of the goal we want to achieve. The gradient is then backpropagated through the Lenia steps to optimize both the parameters of the rule $\theta_l$ and the initialization $A_l$ (details in appendix \ref{appendix:gradient}). As stated before, the obstacles are placed only on one side of a 256x256 grid. In total at every rollout 8 disk of radius 10 are randomly placed as obstacles.

Note that we filter from the history parameters leading to a collapse (mass reaches 0) and explosion of the pattern (pattern expanding too much ) both when initializing the history with random parameters and also after an optimization loop (when the optimization fails) so that we do not use them as starting point for optimization in next steps. More details on the filter we applied can be found in appendix \ref{appendix:history_reuse_filter}.

As presented before, our IMGEP outputs 160 parameters for each seed: 40 from the initialization of history and 120 from the IMGEP steps afterward (1 for each step). We discard the intermediate result of optimization and in each step of the IMGEP only save the final result of the optimization.

The initialization of the history plays an important role in the subsequent steps of the methods as all the following steps will be built on top of this basis, see Fig \ref{fig:curriculum}.a. We thus introduce an initialization selection in order to find promising initialization of the history. More details on this initialization selection mechanism can be found in appendix \ref{appendix:imgep_init_selection}. Note that those steps are counted in the total number of lenia rollouts performed by the method for a fair comparison with random search, and are the main source of stochasticity in the number of rollout performed by a run of the method.

\subsection*{Robustness Evaluation} \label{matmethods:tests}To measure the robustness of the agent against obstacles in the ``basic obstacle test'', we run 50 rollouts of 2000 timesteps with different obstacles positions. Each rollout environment has 23 obstacles of radius 10 randomly sampled uniformly in the whole grid and one placed in the trajectory of the moving agent (to be sure that it encounters obstacles), more details in appendix \ref{appendix:basic_osbtacle_test}. At the end of the 2000 timesteps, we compute statistics on the system rollout to detect if the matter is considered as an agent. We refer to appendix \ref{appendix:agency_test} for more information on the statistics used for empirical agency and robustness tests. We then compute the ratio between the number of rollouts(ie environments) where the pattern survived (passed the empirical agency test) and the total number of rollouts. Robustness is measured similarly in the generalization tests but with 10 rollouts instead of 50. See appendix \ref{appendix:generalization_tests} for more information on the different generalization tests.

\subsection*{Handmade search} 
The parameters from the original lenia papers \cite{chan2019lenia,chan2020lenia} are obtained from: \href{https://github.com/Chakazul/Lenia}{\tiny{https://github.com/Chakazul/Lenia}}. We filter out the ones with multiple channels and the ones with an initialization that does not fit in the 256x256 grid, more details in appendix \ref{appendix:handmade}.

\subsection*{Code and data availability} 
 Code to reproduce the experiments and results can be found on Github at \href{https://github.com/flowersteam/sensorimotor-lenia-search}{\tiny{https://github.com/flowersteam/sensorimotor-lenia-search}}. We also provide the resulting parameters as well as their measured performance on the test tasks in the data folder of the github repository (More details in SI Appendix). An interactive demo can be found at \href{https://developmentalsystems.org/sensorimotor-lenia-companion/}{\tiny{https://developmentalsystems.org/sensorimotor-lenia-companion/}}.

}

\showmatmethods{} 

\acknow{This research was partially funded by the French National Research Agency (\url{https://anr.fr/}), project ECOCURL Grant ANR-20-CE23-0006 and project DeepCuriosity AI chair project. M.E acknowledge support from the biotechnology company Poietis. This work benefited from the use of the Jean Zay supercomputer associated with the Genci grant A0091011996.}

\showacknow{} 

\bibliography{sensori_main}

\section{Appendix}

\begin{itemize}

\item In the first part of this appendix we provide several additional results : \begin{itemize}
    \item In section \ref{appendix:Phylogeny}, we provide the resulting curriculum ``phylogeny'' from a run of IMGEP.
    \item In section \ref{appendix:ablation}, we provide ablation of the IMGEP method: removing obstacles from the training in \ref{appendix:imgep_no_obstacles}, replacing the gradient with a simple evolutionary algorithm in \ref{appendix:imgep_nograd}, and replacing the biased goal sampling by an uniform goal sampling in \ref{appendix:imgep_randomsampling}.
    \item In section \ref{appendix:seed_var}, we provide results for each of the 10 seeds to display the variability. 
    \item In section \ref{appendix:generalization_tests_results}, we provide the full results for the generalization tests. 
\end{itemize}

\item We then provide the details of the method, system and tests :
\begin{itemize}
    \item In section \ref{appendix:lenia_system}, we describe the Lenia system in details. In particular in subsection \ref{appendix:lenia_diff}, we describe the change made on the original lenia system from \cite{chan2019lenia,chan2020lenia} to make it more differentiable.
    \item In section \ref{appendix:imgep_details}, we describe the IMGEP method in details. 
    \item In section \ref{appendix:tests}, we provide details about the tests and measures used in the main papers: empirical agency test in \ref{appendix:agency_test}, moving test in \ref{appendix:moving_test}, speed measure in \ref{appendix:speed_measure}, basic obstacle test in \ref{appendix:basic_osbtacle_test}, generalization tests in \ref{appendix:generalization_tests}.
    \item In section \ref{appendix:comparison_baselines}, we provide details about the baselines we use for comparison: random search in \ref{appendix:random_search}, agent from the original lenia papers \cite{chan2019lenia,chan2020lenia} in \ref{appendix:handmade}
\end{itemize}

\item We provied in section \ref{appendix:legend_videos} the legends of the movies.
\end{itemize}

\subsection{Data availability}

The resulting parameters as well as their measured performances on the tests tasks are available on Github at \url{https://github.com/flowersteam/sensorimotor-lenia-search} in the data folder. More precisely: 
\begin{itemize}
    \item Folder $imgep\_exploration$ contains parameters generated by the IMGEP method presented in the main text as well as their measured robustness.
    \item Folder $random\_exploration$ contains parameters generated by random exploration as well as their measured robustness.
    \item Folder $handmade\_exploration$ contains parameters from the original Lenia papers \cite{chan2019lenia,chan2020lenia} (more details in appendix \ref{appendix:handmade}) as well as their measured robustness.
    \item Folder $imgep\_no\_grad\_init\_exploration$ contains parameters obtained from the IMGEP with ablation on the gradient (described in appendix \ref{appendix:imgep_nograd}) as well as their measured robustness.
    \item Folder $imgep\_no\_obstacles\_exploration$ contains parameters obtained from the IMGEP with ablation of the obstacles (described in appendix\ref{appendix:imgep_no_obstacles}) as well as their measured robustness.
    \item Folder $imgep\_random\_sample\_init\_exploration$ contains parameters obtained from the IMGEP with a uniform sampling of goals (described in appendix \ref{appendix:imgep_randomsampling}) as well as their measured robustness.
    \item Folder $videos$ contains all video presented in this work.
    \item File $creatures\_categories.json$ contains the result of the agency and moving test for all the pre-filtered parameters (more details on the pre-filter in appendix \ref{appendix:agency_test}) from the IMGEP, random, handmade exploration and "IMGEP no obstacles".
    \item File $creatures\_categories\_ablation.json$ contains the result of the agency and moving test for all the pre-filtered parameters (more details on the pre-filter in appendix \ref{appendix:agency_test}) from the ablations presented in appendix \ref{appendix:imgep_nograd} and \ref{appendix:imgep_randomsampling}. 
    
\end{itemize}

We also provide the code to reproduce the experiments on Github at \url{https://github.com/flowersteam/sensorimotor-lenia-search}.

\subsection{Curriculum phylogeny}
\label{appendix:Phylogeny}

\begin{figure}
\centering
\includegraphics[width=.9\linewidth]{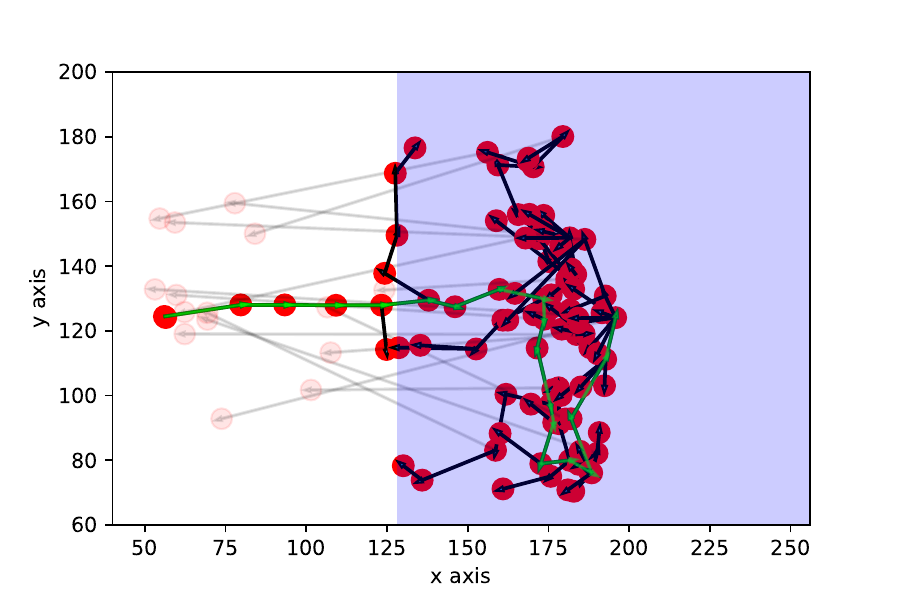}
\caption{``Phylogeny tree'' of one run of IMGEP. The red dot are reached positions (by a step of IMGEP). The blue zone correspond to the zone where obstacles can be placed. Black arrows indicate optimization progress (the point at the end of the arrow was obtained after optimizing the one at the start of the arrow). The path leading to the best agent (reaching the furthest position on the x axis) is highlighted in green. Interestingly we can see that the best path is not necessarily a straight path. For visibility reasons, we put transparency on the optimization steps that led to reached positions far from the reached position of the parameters that was used to initialize the optimization (often due to failing ).}
\label{fig:phylogeny}
\end{figure}

In Fig.\ref{fig:phylogeny}, we explore the curriculum path that is generated by the IMGEP. For this aim, we plot the achieved position (reached goal) by each step of the IMGEP. Arrows show, for each step, what was the previous step result used as initialization. In addition, we highlight in green the sequence of reached positions leading to the furthest position attained. We observe that the path to this furthest position is far from being straightforward. This indicates a rather complex optimization landscape toward this position, that would have been difficult to navigate through gradient descent alone. By generating diverse goals and their associated solutions in parameter space, the IMGEP is able to explore potential stepping stones that can later on prove useful to reach difficult positions.

\subsection{Ablations} \label{appendix:ablation}

We will call the training procedure described in the main text as the \textit{original method}, to which we provide additional detail in \ref{appendix:imgep_details}. In this section, we provide ablation studies aiming to evaluate the effect of removing different components of this original method. To make it as fair as possible and also highlight the difference each ablation introduces, all ablation studies except the ``IMGEP no obstacle'' were made starting with the same initialization of the history as the ones obtained from the initialization search (\ref{appendix:imgep_init_selection}) of the original method. This initialization might however be influenced by the presence of obstacles, this is why ``IMGEP no obstacle'' will run its own initialization search.

\begin{figure*}[t!]
\centering
\includegraphics[width=15cm]{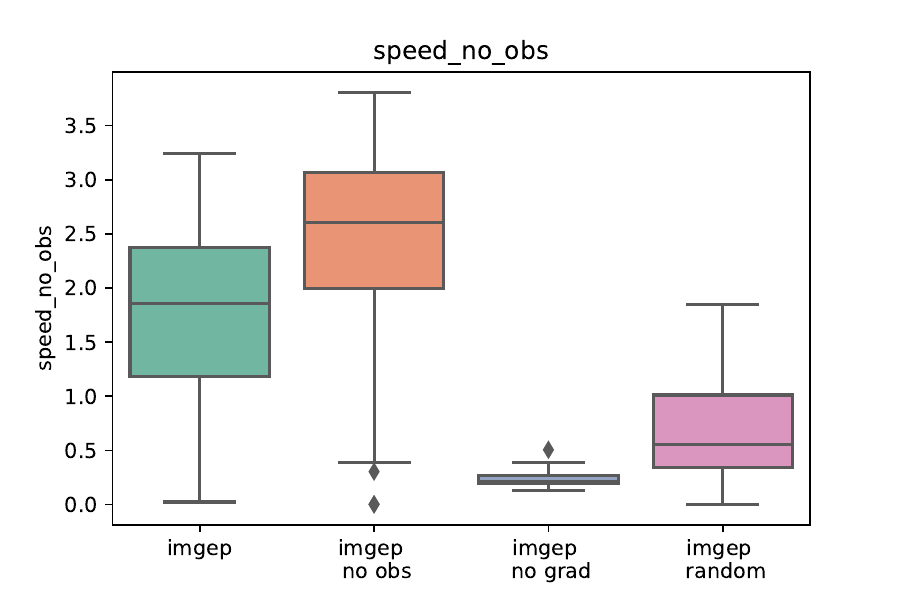}
\caption{Comparison of ablation on speed }
\label{fig:ablation_speed}
\end{figure*}

\begin{figure*}[t!]
\centering
\includegraphics[width=15cm]{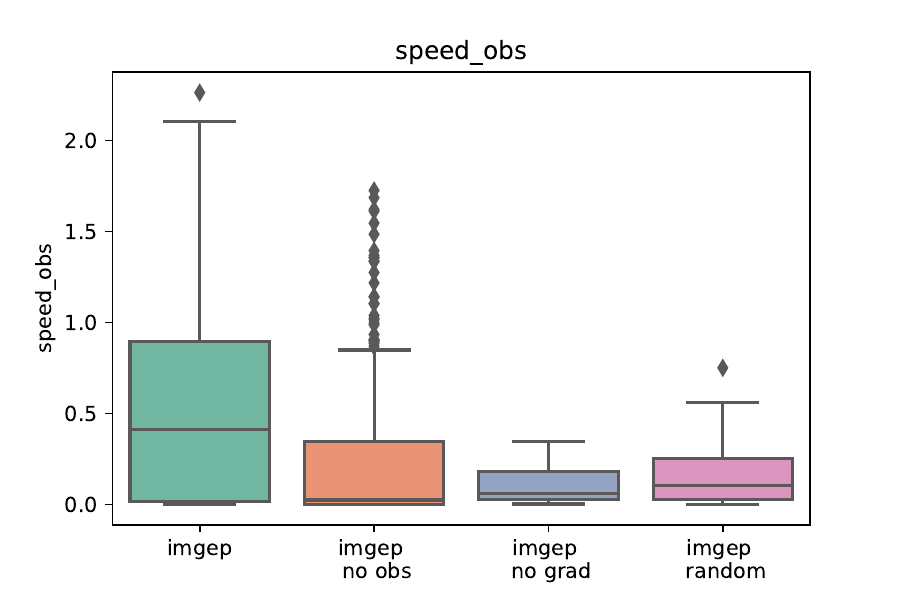}
\caption{Comparison of ablation on speed with obstacles}
\label{fig:ablation_speed_obs}
\end{figure*}

\begin{figure*}[t!]
\centering
\includegraphics[width=15cm]{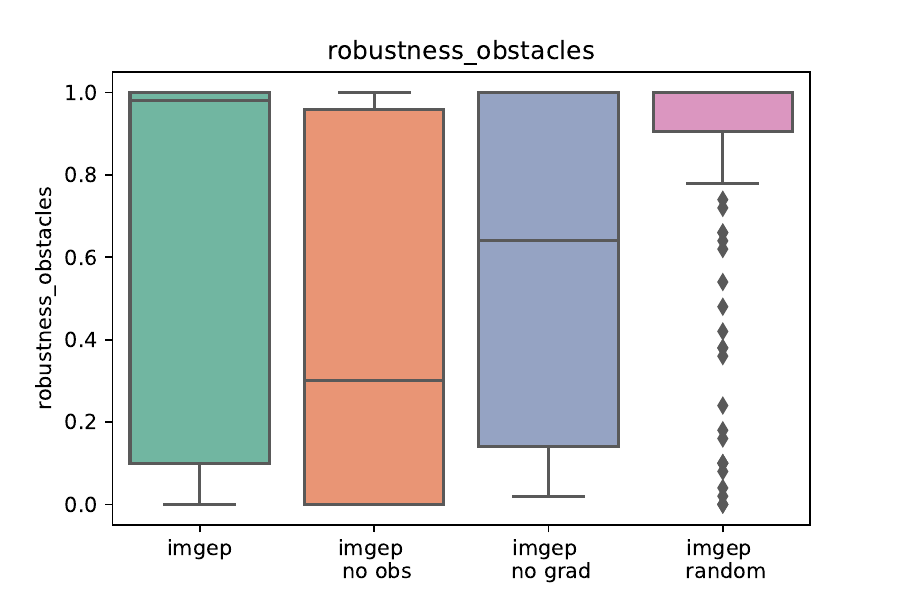}
\caption{Comparison of ablation on robustness to static obstacles}
\label{fig:ablation_robu}
\end{figure*}

\begin{figure*}[t!]
\centering
\includegraphics[width=15cm]{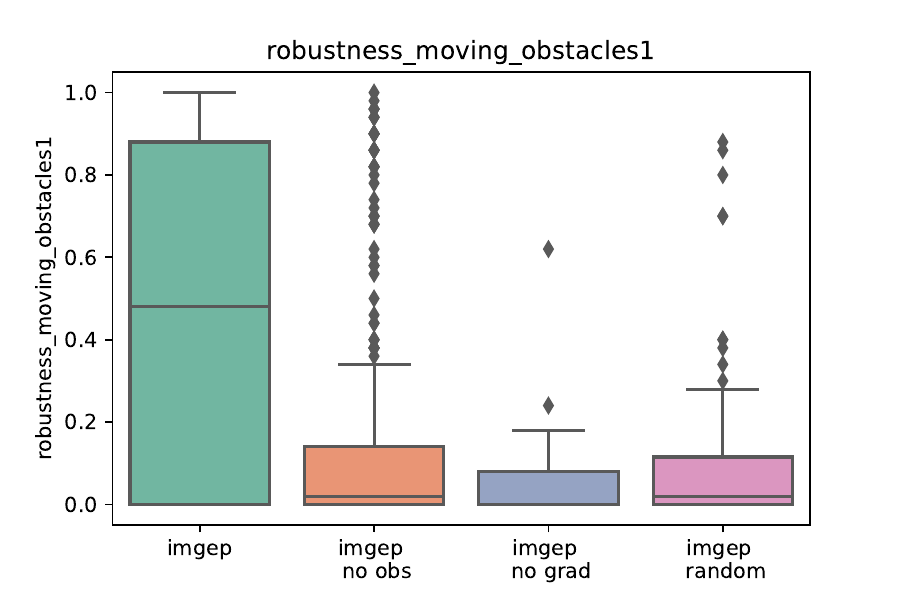}
\caption{Comparison of ablation on robustness to moving obstacles of speed 1}
\label{fig:ablation_robu_moving1}
\end{figure*}

\begin{figure*}[t!]
\centering
\includegraphics[width=15cm]{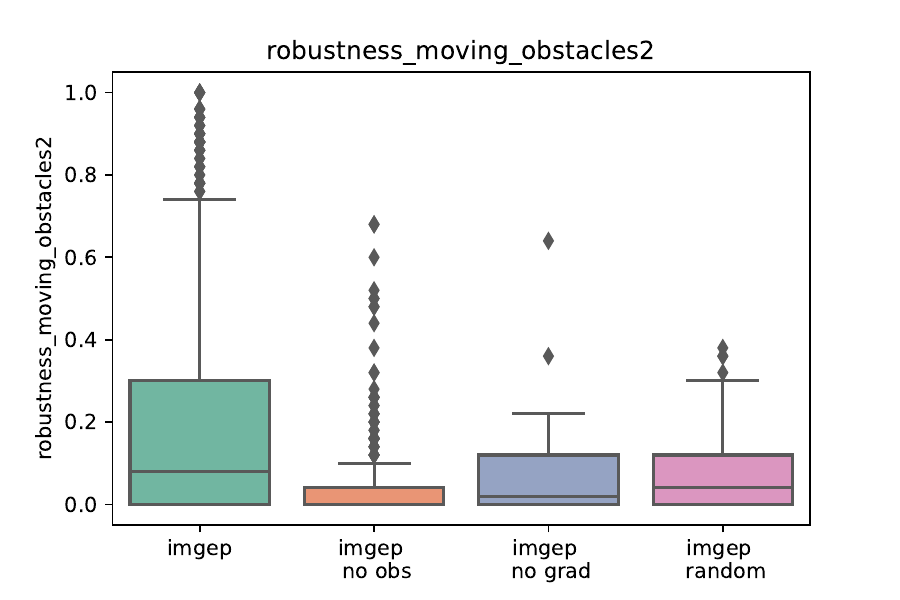}
\caption{Comparison of ablation on robustness to moving obstacles of speed 2}
\label{fig:ablation_robu_moving2}
\end{figure*}

\begin{figure*}[t!]
\centering
\includegraphics[width=15cm]{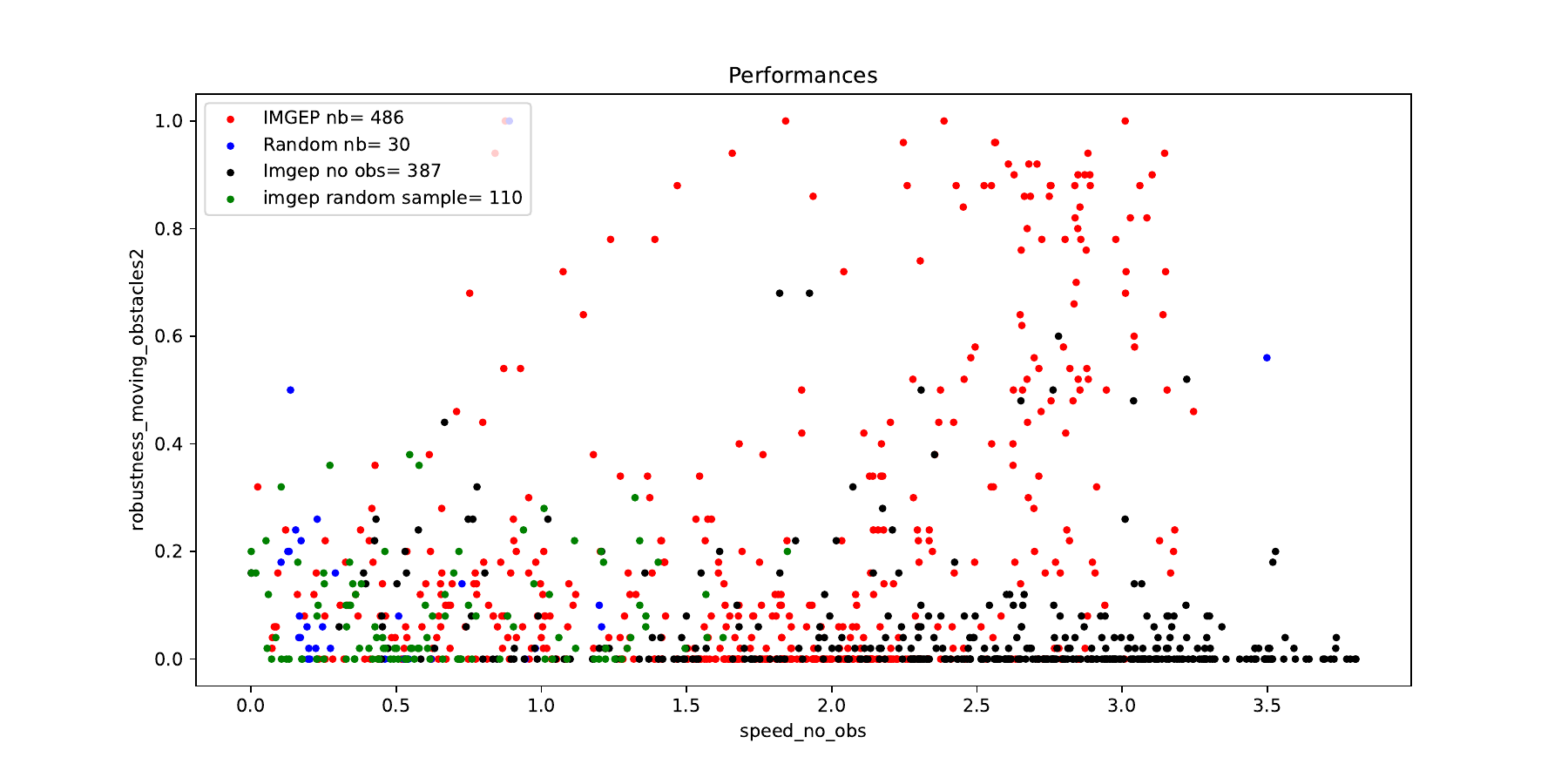}
\caption{Scatter plot of robustness to moving obstacles of speed 2 (y) and speed without obstacles (x) of IMGEP (red), IMGEP without obstacles in the search (black), Random search (blue) and handmade search from the original lenia papers(green). Even for moving agents with comparable speed without obstacles, IMGEP with no obstacles has far less robustness to obstacles of speed 2 than IMGEP trained with obstacles..}
\label{fig:speed_obs_2}
\end{figure*}

\subsubsection{IMGEP no obstacles}

\label{appendix:imgep_no_obstacles} In this ablation, we use the same training procedure as in the original method but remove the obstacles from the grid. This means that during training the agent will only be trained to go further but will never encounter any obstacle.

With this ablation, we obtain moving agents that are faster without obstacles than the original method (Fig. \ref{fig:ablation_speed}) but have far less robustness to obstacles (Fig. \ref{fig:ablation_robu}) and especially here against moving obstacles (Fig. \ref{fig:ablation_robu_moving1},\ref{fig:ablation_robu_moving2}). We also observe that agents trained in the original condition, at equal speed, are more robust to obstacles than those in this ablation (Fig.\ref{fig:speed_obs_2}). This is intuitive as the training without obstacles facilitates reaching further positions (as there is no obstacle in the grid), resulting in higher speed since the episode duration remains constant. However as they are not optimized to resist obstacles, we observe much lower robustness. 


\begin{figure*}[t!]
\centering
\includegraphics[width=15cm]{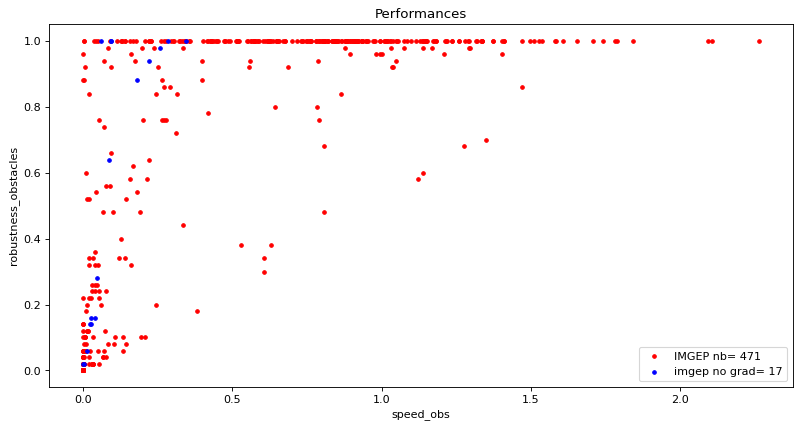}
\caption{Comparison of the original IMGEP method and an IMGEP where gradient descent optimization of the parameters is replaced by random mutations as described in \ref{appendix:imgep_nograd}. We can see that random mutation hardly succeed in optimizing the parameters leading to very poor performance compared to the IMGEP with gradient descent.}
\label{fig:scatter_imgep_nograd}
\end{figure*}

\begin{figure*}[t!]
\centering
\includegraphics[width=15cm]{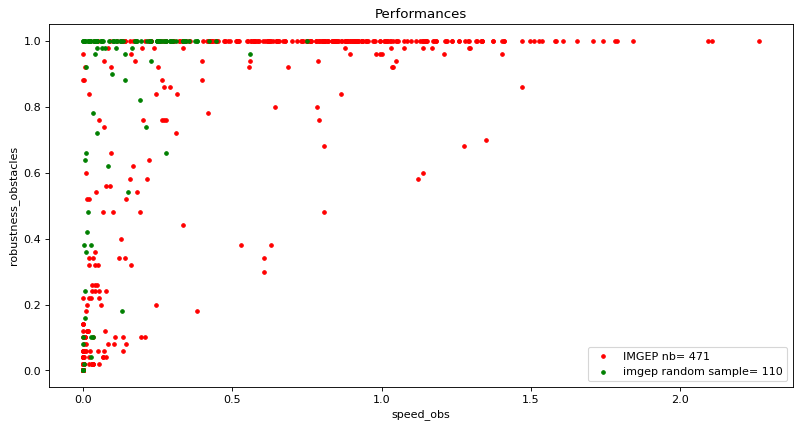}
\caption{Comparison of the original IMGEP method and an IMGEP where our our biased goal sampling is replaced by a random sampling of goal in the grid as described in \ref{appendix:imgep_randomsampling}. We can see that our biased sampling is much more efficient at finding robust fast moving agents.}
\label{appendix_fig:scatter_imgep_randomsample}
\end{figure*}

\subsubsection{No gradient}\label{appendix:imgep_nograd}

In this experiment, we replace the gradient descent in the original method by a simple evolutionary strategy. For each goal we replace the gradient descent by several parallels trials of random mutation (mutation as described in \ref{appendix:mutation_details}) from the candidate parameters with a number of trials equal to the number of gradient descent steps performed during optimization in the original method. At the end of those trials we select the parameters having the lowest loss regarding the goal (same loss as the one used for gradient descent in the original method). We observe that the performances of this method is significantly lower (Fig.\ref{fig:scatter_imgep_nograd}), suggesting that random mutations is not effective in such hard optimization landscapes (and especially with such little number of rollouts) and leads in most cases to explosion or vanish of the matter. 
    
\subsubsection{Uniform Random sampling of target in IMGEP} \label{appendix:imgep_randomsampling}

    \begin{figure*}[t!]
    \centering
    \includegraphics[width=0.49\textwidth]{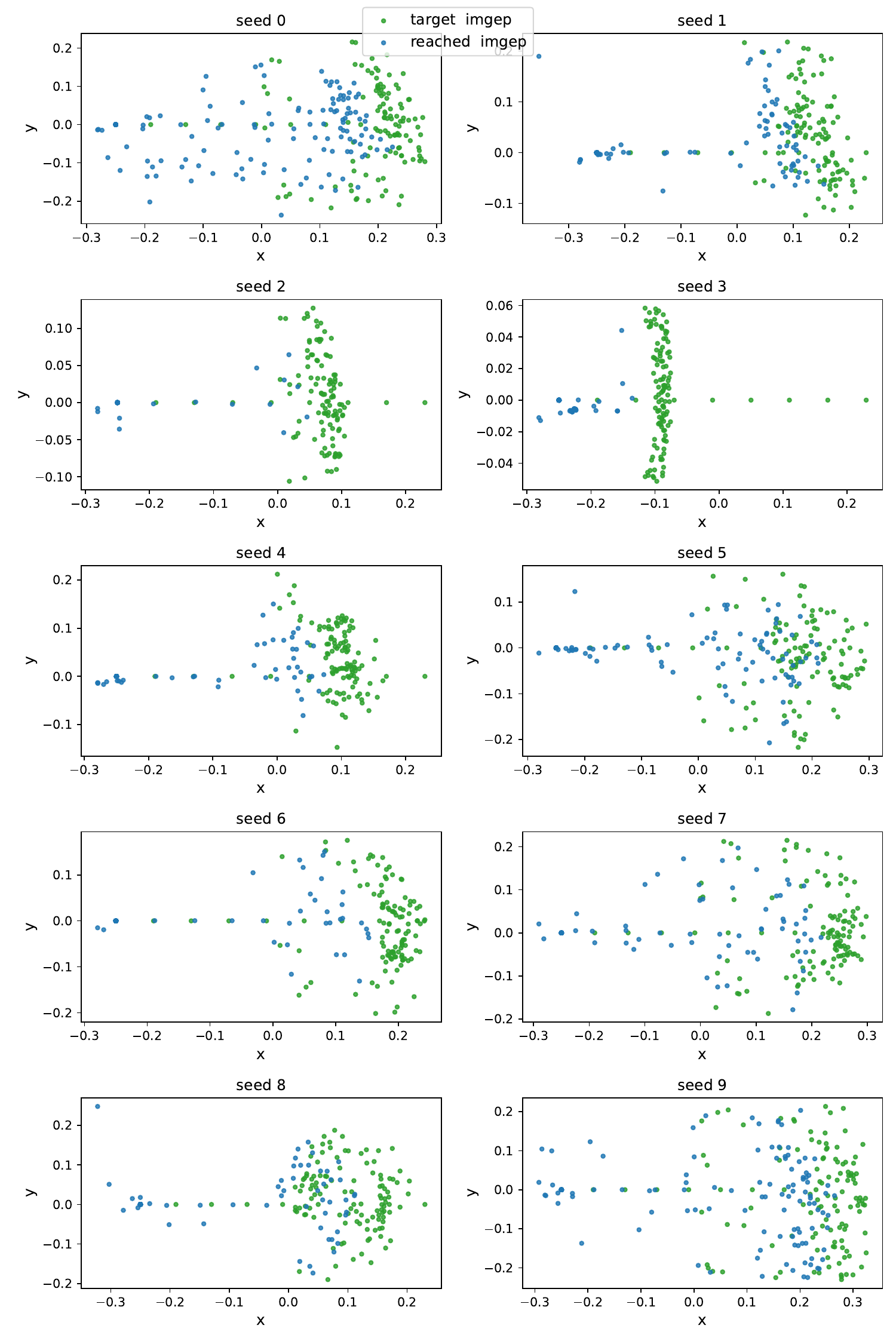}
    \includegraphics[width=0.49\textwidth]{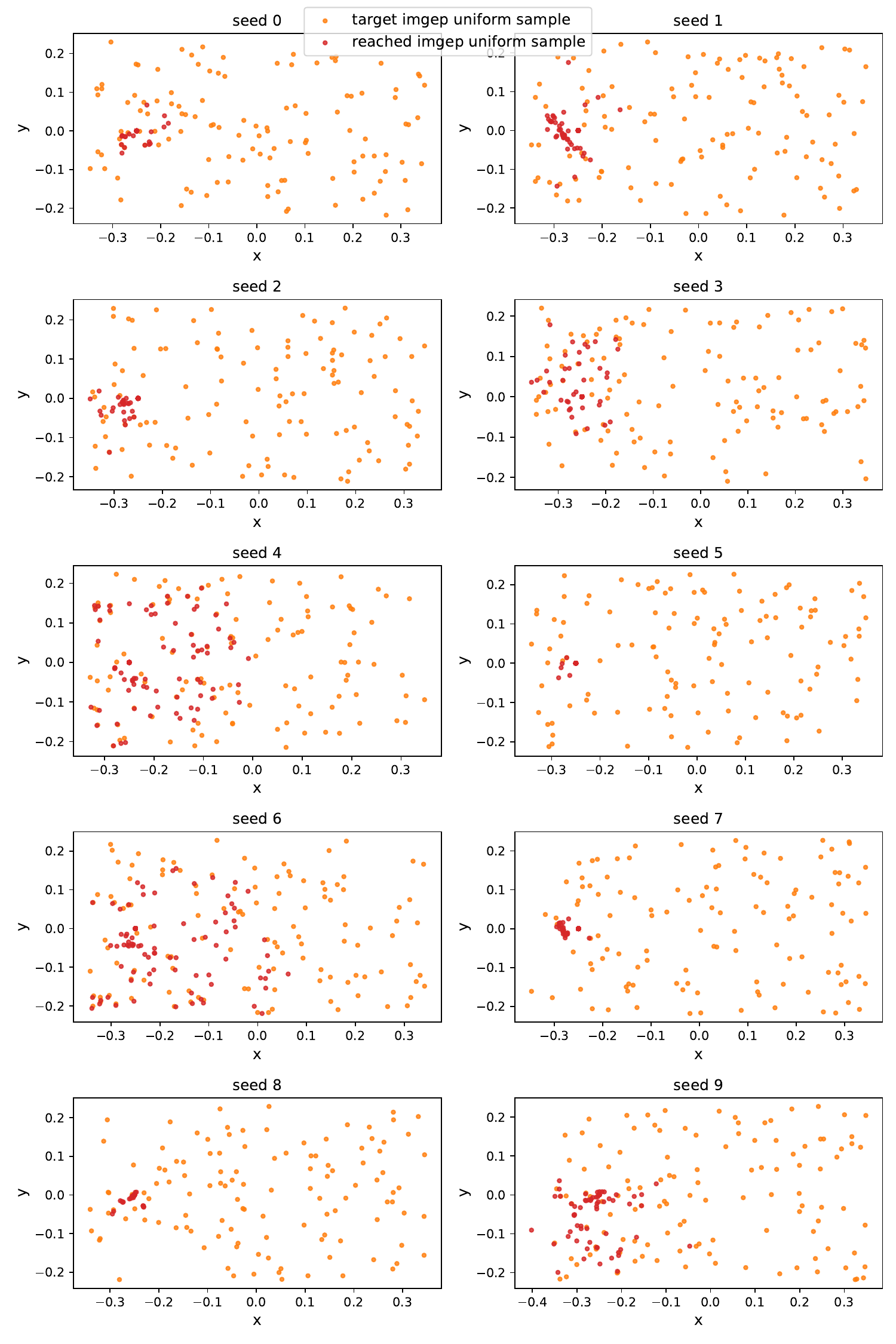}
    \caption{ Target goals and reached positions for every seed of (left) original method (right) IMGEP with uniform sampling of goal. The uniform sampling IMGEP sample a lot of far points that not reached at all}
    \label{fig:target_reached_imgep_random}
    \end{figure*}
    
     In this experiment, we replace the curriculum-driven goal sampling of the original method (detail on curriculum in \ref{appendix:goal_sampling}) by a uniform sampling in the grid.

    Compared to the original method, we observe overall lower performances in term of speed and robustness (Fig. \ref{appendix_fig:scatter_imgep_randomsample} and Fig.\ref{fig:ablation_speed},\ref{fig:ablation_speed_obs},\ref{fig:ablation_robu_moving1},\ref{fig:ablation_robu_moving2}). This can be explained by the fact that random sampling often sample goals that are impossible to reach at the time. We observe that, with the same budget as the original method run, it only reaches a very small subset of the entire grid compared to the original method (Fig.\ref{fig:target_reached_imgep_random}). Most target goals far from initialization failed while goals that were close enough were sometimes successful. 

    However, we observe that this ablation still allows to obtain more moving agents than random search (110 vs. 30)

    We introduced a curriculum in our original method mostly to speed up computation. We indeed show with this ablation how it benefits the search process. Note however that, in theory, the current ablation should obtain similar results if given enough compute budget (but will most likely require much more time). In fact, a curriculum can also emerge with random goal sampling, as the agent will only make progress on goals that either not too far or too close from its current abilities. (see e.g. Forestier et al. 2022 \cite{Forestier2017IntrinsicallyMG}).

\subsection{Seed variability}\label{appendix:seed_var}

We report in tab.\ref{appendix:seed_var_tab} the variability of the results of the method across the 10 seeds. The variability in result might indicate that some parameter area are easier to navigate or more prone to certain behavior. Overall we still observe that every seed finds a good amount of moving agents and most of them find at least 1 robust agent(ie an agent with a score >0.95 to the ``basic obstacle test'').

\begin{table}

\caption{Seed variability}

\footnotesize
\begin{tabular*}{\hsize}{l|ccccc}

Seed&Number&Number&Number&max&max\cr

Number& of agents& of moving&of robust&speed&speed obs\cr
&&(agents)&(agents)&(agents)&(agents)\cr

\hline

Seed 0&107&93&91&2.8&1.4\cr
Seed 1&64&54&26&2.7&1.5\cr
Seed 2&33&32&1&2.0&1.1\cr
Seed 3&18&7&0&0.5&0.3\cr
Seed 4&35&26&6&1.9&0.4\cr
Seed 5&66&52&38&2.9&1.8\cr
Seed 6&54&54&2&2.5&0.3\cr
Seed 7&30&30&1&3.0&0.9\cr
Seed 8&44&44&4&2.3&0.3\cr
Seed 9&104&94&92&3.2&2.3\cr

\hline
\end{tabular*}
\label{appendix:seed_var_tab}
\end{table}

\subsection{Generalization table} \label{appendix:generalization_tests_results}

We refer to table \ref{appendix:generalization_table} for the full generalization results. 

\begin{table}

\caption{Generalization results}

\footnotesize
\begin{tabular*}{\hsize}{lcr cc c c}

\multicolumn{3}{l}{Tests}&
\multicolumn{2}{c}{IMGEP}&
Random &
Handmade \cr

 & & & speed>1 & 10 best & 10 best & 10 best \cr
 
\hline
\multicolumn{3}{l}{speed} & 1.33 $\pm$ 0.28 & 1.94 $\pm$ 0.15 & 0.53 $\pm$ 0.25 & 0.34 $\pm$ 0.10 \cr
\hline
\multicolumn{3}{l}{obstacle} & & & & \cr
\multicolumn{3}{c}{number} & & & & \cr
\multicolumn{3}{r}{24} & 0.98 $\pm$ 0.07 & 0.99 $\pm$ 0.03  & 0.99 $\pm$ 0.03 & 0.99 $\pm$ 0.03 \cr
\multicolumn{3}{r}{30} & 0.98 $\pm$ 0.07 & 1.00 $\pm$ 0.00  & 0.99 $\pm$ 0.03 & 0.99 $\pm$ 0.03 \cr
\multicolumn{3}{r}{36} & 0.99 $\pm$ 0.06 & 1.00 $\pm$ 0.00  & 0.99 $\pm$ 0.03 & 0.97 $\pm$ 0.09 \cr
\multicolumn{3}{r}{42} & 0.99 $\pm$ 0.03 & 1.00 $\pm$ 0.00  & 0.99 $\pm$ 0.03 & 0.97 $\pm$ 0.09 \cr
\multicolumn{3}{r}{48} & 0.99 $\pm$ 0.04 & 1.00 $\pm$ 0.00  & 1.00 $\pm$ 0.00 & 0.98 $\pm$ 0.06 \cr
\multicolumn{3}{c}{radius} & & & & \cr
\multicolumn{3}{r}{4} & 0.92 $\pm$ 0.18 & 0.90 $\pm$ 0.13  & 0.92 $\pm$ 0.12 & 0.95 $\pm$ 0.09 \cr
\multicolumn{3}{r}{7} & 0.98 $\pm$ 0.08 & 1.00 $\pm$ 0.00  & 1.00 $\pm$ 0.00 & 0.97 $\pm$ 0.09 \cr
\multicolumn{3}{r}{10} & 0.98 $\pm$ 0.07 & 0.99 $\pm$ 0.03  & 0.99 $\pm$ 0.03 & 0.99 $\pm$ 0.03 \cr
\multicolumn{3}{r}{13} & 0.98 $\pm$ 0.08 & 0.99 $\pm$ 0.03  & 1.00 $\pm$ 0.00 & 0.99 $\pm$ 0.03 \cr
\multicolumn{3}{r}{16} & 0.98 $\pm$ 0.08 & 1.00 $\pm$ 0.00  & 1.00 $\pm$ 0.00 & 1.00 $\pm$ 0.00 \cr
\multicolumn{3}{c}{speed} & & & & \cr
\multicolumn{3}{r}{1/3} & 0.99 $\pm$ 0.04 & 1.00 $\pm$ 0.00  & 0.77 $\pm$ 0.27 & 0.74 $\pm$ 0.28 \cr
\multicolumn{3}{r}{1/2} & 0.97 $\pm$ 0.07 & 1.00 $\pm$ 0.00  & 0.61 $\pm$ 0.38 & 0.51 $\pm$ 0.38 \cr
\multicolumn{3}{r}{1} & 0.81 $\pm$ 0.23 & 0.97 $\pm$ 0.05  & 0.42 $\pm$ 0.41 & 0.02 $\pm$ 0.04 \cr
\multicolumn{3}{r}{2} & 0.34 $\pm$ 0.32 & 0.71 $\pm$ 0.25  & 0.13 $\pm$ 0.29 & 0.00 $\pm$ 0.00 \cr
\multicolumn{3}{r}{3} & 0.12 $\pm$ 0.15 & 0.32 $\pm$ 0.17  & 0.07 $\pm$ 0.12 & 0.00 $\pm$ 0.00 \cr
\hline
\multicolumn{3}{l}{update} & & & & \cr
\multicolumn{3}{c}{mask rate} & & & & \cr
\multicolumn{3}{r}{0.2} & 0.99 $\pm$ 0.08 & 1.00 $\pm$ 0.00  & 1.00 $\pm$ 0.00 & 1.00 $\pm$ 0.00 \cr
\multicolumn{3}{r}{0.6} & 0.99 $\pm$ 0.08 & 1.00 $\pm$ 0.00  & 0.89 $\pm$ 0.30 & 1.00 $\pm$ 0.00 \cr
\multicolumn{3}{r}{1.0} & 1.00 $\pm$ 0.00 & 1.00 $\pm$ 0.00  & 1.00 $\pm$ 0.00 & 1.00 $\pm$ 0.00 \cr
\multicolumn{3}{r}{1.4} & 0.99 $\pm$ 0.09 & 1.00 $\pm$ 0.00  & 1.00 $\pm$ 0.00 & 1.00 $\pm$ 0.00 \cr
\multicolumn{3}{r}{1.8} & 0.99 $\pm$ 0.10 & 1.00 $\pm$ 0.00  & 1.00 $\pm$ 0.00 & 1.00 $\pm$ 0.00 \cr
\multicolumn{3}{c}{noise rate} & & & & \cr
\multicolumn{3}{r}{0.2} & 0.91 $\pm$ 0.28 & 0.90 $\pm$ 0.30  & 0.77 $\pm$ 0.37 & 0.99 $\pm$ 0.03 \cr
\multicolumn{3}{r}{0.4} & 0.75 $\pm$ 0.42 & 0.91 $\pm$ 0.27  & 0.74 $\pm$ 0.38 & 0.92 $\pm$ 0.18 \cr
\multicolumn{3}{r}{0.6} & 0.67 $\pm$ 0.45 & 0.90 $\pm$ 0.27  & 0.58 $\pm$ 0.46 & 0.77 $\pm$ 0.38 \cr
\multicolumn{3}{r}{0.8} & 0.60 $\pm$ 0.47 & 0.63 $\pm$ 0.44  & 0.50 $\pm$ 0.44 & 0.71 $\pm$ 0.44 \cr
\multicolumn{3}{r}{1.0} & 0.51 $\pm$ 0.47 & 0.32 $\pm$ 0.41  & 0.44 $\pm$ 0.45 & 0.70 $\pm$ 0.46 \cr
\multicolumn{3}{c}{noise std} & & & & \cr
\multicolumn{3}{r}{0.2} & 0.99 $\pm$ 0.11 & 1.00 $\pm$ 0.00  & 0.96 $\pm$ 0.12 & 1.00 $\pm$ 0.00 \cr
\multicolumn{3}{r}{0.6} & 0.79 $\pm$ 0.39 & 0.90 $\pm$ 0.30  & 0.76 $\pm$ 0.39 & 0.98 $\pm$ 0.06 \cr
\multicolumn{3}{r}{1.0} & 0.51 $\pm$ 0.47 & 0.32 $\pm$ 0.41  & 0.44 $\pm$ 0.45 & 0.70 $\pm$ 0.46 \cr
\multicolumn{3}{r}{1.4} & 0.08 $\pm$ 0.21 & 0.03 $\pm$ 0.09  & 0.18 $\pm$ 0.32 & 0.56 $\pm$ 0.45 \cr
\multicolumn{3}{r}{1.8} & 0.06 $\pm$ 0.14 & 0.06 $\pm$ 0.10  & 0.17 $\pm$ 0.30 & 0.45 $\pm$ 0.47 \cr
\hline
\multicolumn{3}{l}{init} & & & & \cr
\multicolumn{3}{c}{noise rate} & & & & \cr
\multicolumn{3}{r}{0.2} & 1.00 $\pm$ 0.01 & 1.00 $\pm$ 0.00  & 0.89 $\pm$ 0.16 & 1.00 $\pm$ 0.00 \cr
\multicolumn{3}{r}{0.4} & 0.99 $\pm$ 0.09 & 1.00 $\pm$ 0.00  & 0.91 $\pm$ 0.24 & 0.99 $\pm$ 0.03 \cr
\multicolumn{3}{r}{0.6} & 0.98 $\pm$ 0.13 & 1.00 $\pm$ 0.00  & 0.88 $\pm$ 0.30 & 0.95 $\pm$ 0.15 \cr
\multicolumn{3}{r}{0.8} & 0.97 $\pm$ 0.14 & 1.00 $\pm$ 0.00  & 0.88 $\pm$ 0.30 & 0.89 $\pm$ 0.24 \cr
\multicolumn{3}{r}{1.0} & 0.95 $\pm$ 0.21 & 1.00 $\pm$ 0.00  & 0.88 $\pm$ 0.30 & 0.76 $\pm$ 0.29 \cr
\multicolumn{3}{c}{noise std} & & & & \cr
\multicolumn{3}{r}{0.5} & 0.97 $\pm$ 0.16 & 1.00 $\pm$ 0.00  & 0.87 $\pm$ 0.30 & 0.97 $\pm$ 0.09 \cr
\multicolumn{3}{r}{1.5} & 0.94 $\pm$ 0.20 & 0.98 $\pm$ 0.06  & 0.85 $\pm$ 0.30 & 0.52 $\pm$ 0.42 \cr
\multicolumn{3}{r}{2.5} & 0.89 $\pm$ 0.27 & 0.92 $\pm$ 0.17  & 0.80 $\pm$ 0.36 & 0.37 $\pm$ 0.44 \cr
\multicolumn{3}{r}{3.5} & 0.86 $\pm$ 0.32 & 0.91 $\pm$ 0.27  & 0.81 $\pm$ 0.34 & 0.35 $\pm$ 0.45 \cr
\multicolumn{3}{r}{4.5} & 0.85 $\pm$ 0.32 & 0.94 $\pm$ 0.18  & 0.79 $\pm$ 0.38 & 0.32 $\pm$ 0.43 \cr
\hline
\multicolumn{3}{l}{scaling} & & & & \cr
\multicolumn{3}{r}{0.15} & 0.91 $\pm$ 0.28 & 0.90 $\pm$ 0.30  & 0.30 $\pm$ 0.46 & 0.00 $\pm$ 0.00 \cr
\multicolumn{3}{r}{0.65} & 0.99 $\pm$ 0.10 & 1.00 $\pm$ 0.00  & 0.50 $\pm$ 0.50 & 1.00 $\pm$ 0.00 \cr
\multicolumn{3}{r}{1.15} & 1.00 $\pm$ 0.00 & 1.00 $\pm$ 0.00  & 0.70 $\pm$ 0.46 & 1.00 $\pm$ 0.00 \cr
\multicolumn{3}{r}{1.65} & 1.00 $\pm$ 0.00 & 1.00 $\pm$ 0.00  & 0.70 $\pm$ 0.46 & 1.00 $\pm$ 0.00 \cr
\multicolumn{3}{r}{2.15} & 1.00 $\pm$ 0.00 & 1.00 $\pm$ 0.00  & 0.60 $\pm$ 0.49 & 1.00 $\pm$ 0.00 \cr
\hline
\end{tabular*}
\label{appendix:generalization_table}
\end{table}

\subsection{Lenia system}\label{appendix:lenia_system}

Cellular automata are, in their classic form, a grid of “cells” $ A = \{ a_x \}$ that evolve through time $A^{t=1}\longrightarrow ... \longrightarrow A^{t=T}$ via local “physics-like” laws. More precisely, the cells sequentially update their state based on the states of their neighbours: $ a_x^{t+1}= f(a_x^t,\mathcal{N}(a_x^t))$, where $ x \in \mathcal{X}$ is the position of the cell on the grid, $a_x $ is the state of the cell, and $\mathcal{N}(a_x^t)$ is the neighbourhood of the cell. The dynamic of the CA is thus entirely defined by the initialization $ A^{t=1} $ (initial state of the cells in the grid) and the update rule $f$ (function that takes a scalar and outputs a scalar,control how a cell updates based on its neighbours). But predicting their long term behavior is a difficult challenge even for simple ones due to their chaotic dynamics.

Lenia is a class of continuous cellular automata (CA) where each CA instance is defined by a set of parameters $\theta$ that conditions the CA rule $f_{\theta}$; once the parameters $\theta$ conditioning the update rule has been chosen, the system is a classical CA where the initial grid pattern $A^{t=1}$ will be updated.

In Lenia, the system is composed of several communicating grids $ A=\{ A_c\}$ which we call channels. In each of these grids, every cell/pixel can take any value between 0 and 1. Cells at 0 are considered dead while others are alive. The channels are updated in parallel according to their own physics rule. Intuitively, we can see channels as the domain of existence of a certain type of cell. Each type of cell has its own physics : it has its own way to interact with other cells of its type (intra-channel influence) and also its own way to interact with cells of other types (cross-channel influence).

The update of a cell $ a_{x,c}$ at position $x$ in channel $c$ can be decomposed in three steps. First the cell senses its neighbourhood in some other channels (its neighbourhood in its channel, with cells of the same type but also in other channels with other types of cells) through convolution kernels which are filters $K_k$ of different shapes and sizes. Second, the cell converts this sensing into an update (whether positive or negative growth or neutral) through growth functions $G_k$ associated with the kernels. Finally, the cell modifies its state by summing the scalars obtained after the growth functions and adding it to its current state. After the update of every rule has been applied, the state is clipped between 0 and 1. Each (kernel,growth function) couple is associated to the source channel $c_s$ it senses, and to the target channel $c_t$ it updates. A couple (kernel, growth function) characterizes a rule on how a type of cell $c_t$ reacts to its neighbourhood of cells of type $c_s$. Note that $c_s$ and $c_t$ could be the same, which correspond to interaction of cells of the same type (intra-channel influence). Note also that we can have several rules, i.e. several (kernel,growth function) couples, characterizing the interaction between $c_s$ and $c_t$. 

A local update in the grid is summarized with the following formula (where $G^k,K^k,c_s^k,c_t^k$ are respectively the growth function, convolution filter, source channel, target channel associated with the k'th rule): 

$$a_x^{t+1}=f(a_x^t, \mathcal{N}(a_x^t)) =$$ $$\begin{bmatrix} a^t_{x,c_0}  + \frac{1}{T}\sum_{k\text{ st } c_t^k=0}  G^k( K^k(a^t_{x,c_s^k}, \mathcal{N}_{c_s^k}(a^t_x))) \\.\\.\\.\\ a^t_{x,c_C}  + \frac{1}{T}\sum_{k\text{ st } c_t^k=C}  G^k( K^k(a^t_{x,c_s^k}, \mathcal{N}_{c_s^k}(a^t_x)))\end{bmatrix} $$

For each rule, the shape of the (kernel, growth function) is parameterized. We are thus able to “tune” the physics of the cells and of their interactions by changing the kernels shape (how the cells perceive their neighborhood) as well as the growth function shape (how the cells react to this perception).

\subsubsection{Differentiating through Lenia steps} \label{appendix:lenia_diff}

Due to the locality and recurrence of the update rule, there is a close relationship between cellular automata and recurrent convolutional networks \cite{gilpin2019cellular}. In fact, we can see a rollout in Lenia as applying a recurrent neural network to an initial state. If (some of) the network parameters are differentiable, backpropagation can be done by “unfolding” the Lenia rollout and applying a loss at certain time step(s) like in \cite{mordvintsev2020growing}.

However, in the classic version of Lenia, the shape of the kernels are not totally differentiable and not very flexible. To allow easier optimization of the Lenia system, we introduce some changes to the kernel parameterization.

\begin{figure}
\centering
\includegraphics[width=.8\linewidth]{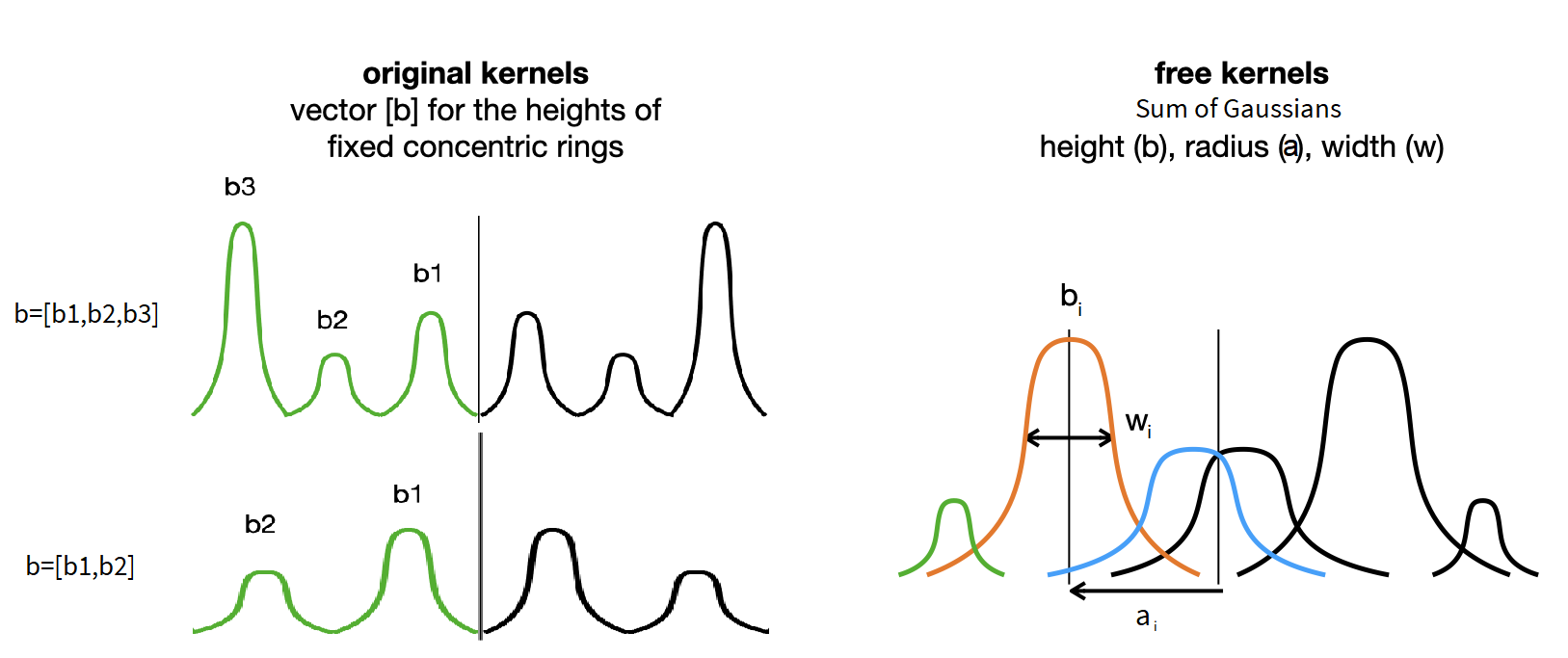}
\caption{Visualization of (left) the convolution kernels used in the original lenia papers \cite{chan2019lenia,chan2020lenia}, (right) the kernel we propose in this paper for more differentiation capabilities. The kernel we propose consists of a sum of free shifted gaussian bumps while the one in the original lenia papers consist of fixed concentrated rings. }
\label{fig:diff_lenia}
\end{figure}

In fact in the original Lenia \cite{chan2019lenia}, the number of bumps in the kernel (see Fig.\ref{fig:diff_lenia} left ) is fixed and cannot be optimized through gradient descent. 

We therefore introduced a new class of CA with differentiable parameters. To do so, the main shift is to use kernels in the form of a sum of k overlapping gaussian bumps: $$x \rightarrow \sum_i^{k} b_i exp(-\frac{(\frac{x}{rR}-a_i)^2}{2w_i^2}) $$ The parameters controlling the shape are then $3 k$-dimensional vectors: $b$ for height of the bump, $w$ for the size of the bump and $a$ for the center of the bump.

These symmetric “free kernels”, while very inspired from Lenia’s original “vanilla bumps”, allow differentiation and more flexibility and expressivity but at the cost of more parameters. For example, it is possible to reduce the number of bumps by assigning some null height values, allowing the number of bumps to be optimized through gradient descent. 

In Lenia, a growth function $ G: [0,1] \rightarrow [-1,1] $ is any unimodal non-monotonic function that satisfies $ G(\mu)=1 $. In this work, we use the continuous exponential growth function $ G(x) = 2 \exp \left( - \frac{(x - \mu )^2}{2 \sigma^2} \right) - 1 $ which is differentiable with respect to $\mu$ and $\sigma$. 

To summarize, the parameters of the update rule are thus those controlling the kernel shape ($R,r,a,w,b$), those controlling the growth function ($\mu,\sigma,h$) and a time controlling parameter ($T$). For a total of $n$ rules (all channels included) with $k$ bumps kernels, the number of parameters is $ (3k+4)n + 2 $. In our experiments, R and T are chosen randomly and fixed while all the other parameters are optimized, and we use a total of $n=10$ rules with $k=3$ bumps kernels . So in total we have 132 parameters for the rules from which 130 are optimized.

In addition to the rule, parameters we also optimize the initialization square $I_{square} \in [0,1]^{(40,40)}$.

\subsubsection{Obstacles}\label{appendix:lenia_obstacles}

The multi-channel aspect of Lenia allows the implementation of different types of cells/particles. To implement obstacles in Lenia we added a separate “obstacle” channel with a kernel going from this channel to the learnable “creature” channel (see Fig.\ref{fig:env_description}). This kernel triggers a severe negative growth in the pixels of the learnable channel where there are obstacles but has no impact on other pixels where there are no obstacles (very localized kernel). This way we prevent any growth in the pixels of the learnable channel where there are obstacles. The formula of the growth function is : $ G(x) = - clip((x-1e-8),0,1)*10 $. Hyperparameters of this handmade rule can be found in \ref{appendix:lenia_params}.



The learnable channel cells can only sense the obstacles through the changes/deformations it implies on it or its neighbours. In fact, as the only kernel that goes from the obstacle channel to the learnable channel is the one we hand-designed, if a macro agent emerges it has to “touch” the obstacle to sense it. To be precise the agent can only sense an obstacle because its interaction with the obstacle will perturb its own configuration and dynamics (i.e. its shape and the interaction between the cells constituting it). This is similar to experiments with swarming bacteria \cite{PhysRevE.101.012407}, where the swarm agent must learn to collectively avoid antibiotic zones (externally-added obstacles) where the bacteria can’t live.

 In our implementation, obstacles stay still, meaning that there is no rule that goes toward (and hence no update of) the obstacle channel . As such, an update step in the final system is summarized at the bottom of Fig.\ref{fig:env_description}.

To test the agents under moving obstacles, we simply shift the channel of obstacles of a certain amount of pixel at every timestep. This shift of the grid, for an integer value of speed, can be written as a rule of the system from the obstacle channel to the obstacle channel. The rule would be the same on all the grid and is localized as it is a function of the fixed neighbourhood. Moving obstacles with a speed with a rational value (for example 0.5 pixels/timesteps) is done in our case by doing the shift every few timesteps.

\subsubsection{Lenia rules parameters}\label{appendix:lenia_params}

Here is the list of the parameters associated to the rules of a Lenia system with C channels, $nb_k$ rules with kernels with k bumps. We also provide the range used in this work for the learnable channel. In this work we used C=2 channels (one learnable channel and the fixed channel), $nb_k=10$ learnable rules and 1 fixed rule (for the obstacles). 
\begin{itemize}
    \item{Common to all rules}
        \begin{itemize}
            \item T $\in [1,10]$ 
        \end{itemize}
    \item{Learnable rules}
    \begin{itemize}
        
        \item \underline{Kernel (convolution filter) parameters:}
            \begin{itemize}
                \item R $\in [15,40]$ Radius of the kernels (common to all kernels)
                \item r $\in [0,1]^{nb_k}$ relative radius of each kernel.
                \item b $\in [0,1]^{nb_k,k}$ height of the k bumps.
                \item w $\in [0.01,0.5]^{nb_k,k}$ width of the k bumps.
                \item a $\in [0,1]^{nb_k,k}$ position of the bumps on the radius. 
            \end{itemize}
        \item \underline{Growth function $ G(x) = 2 \exp \left( - \frac{(x - \mu )^2}{2 \sigma^2} \right) - 1 $ parameters}
        
            \begin{itemize}
                \item $\mu \in [0.05,0.5]^{nb_k}$ mean of the gaussian growth function.
                \item $\sigma \in [0.001,0.18]^{nb_k}$ variance of the gaussian growth function.
                \item h $\in [0,1]^{nb_k}$
            \end{itemize} 
        \item $c_0 = [0]\times nb_k$ source channel (0 is learnable channel)
        \item $c_1 = [0]\times nb_k$ destination channel 
    \end{itemize}

    \item{Fixed rule}
    \begin{itemize}
        \item \underline{Kernel parameters:}
            \begin{itemize}
                \item R = 4 small radius for very localized action
                \item r = [1,1,1]
                \item b = [1,0,0]
                \item w = [0.5,1,1]
                \item a = [0,0,0] 
            \end{itemize}
        \item \underline{Growth function $ G(x) = - clip((x-1e-8),0,1)*10  $ }
        \item $c_0 = 1$ source channel (1 is fixed channel)
        \item $c_1 = 0$ destination channel 
    \end{itemize}
\end{itemize}

\subsubsection{Lenia rule parameter mutations}\label{appendix:lenia_mutation_range}

\begin{itemize}
    \item{Common to all rules}
        \begin{itemize}
            \item T : $\mathcal{N}(0,0.1)\times \mathcal{B}(0.01)$  (mutation then integer)
        \end{itemize}
    \item{Learnable rules}
    \begin{itemize}
        
        \item \underline{Kernel (convolution filter) parameters:}
            \begin{itemize}
                \item R $\mathcal{N}(0,0.1)\times \mathcal{B}(0.01)$ (mutation then integer)
                \item r :$\mathcal{N}(0_{nb_k},0.2 \times \mathcal{I}_{nb_k})$
                \item b : $\mathcal{N}(0_{3nb_k},0.2 \times \mathcal{I}_{3nb_k})$
                \item w :$\mathcal{N}(0_{3nb_k},0.2 \times \mathcal{I}_{3nb_k})$
                \item a : $\mathcal{N}(0_{3nb_k},0.2 \times \mathcal{I}_{3nb_k})$
            \end{itemize}
        \item \underline{Growth function $ G(x) = 2 \exp \left( - \frac{(x - \mu )^2}{2 \sigma^2} \right) - 1 $ parameters}
        
            \begin{itemize}
                \item $\mu$ : $\mathcal{N}(0_{nb_k},0.2 \times \mathcal{I}_{nb_k}) \times \mathcal{B}(0.1)$
                \item $\sigma$: $\mathcal{N}(0_{nb_k},0.01 \times \mathcal{I}_{nb_k}) \times \mathcal{B}(0.1)$
                \item h $\mathcal{N}(0_{nb_k},0.2 \times \mathcal{I}_{nb_k}) \times \mathcal{B}(0.1)$
            \end{itemize}  
    \end{itemize}

\end{itemize}

\subsection{IMGEP details}\label{appendix:imgep_details}

\begin{figure*}[t!]
\centering
\includegraphics[width=15cm]{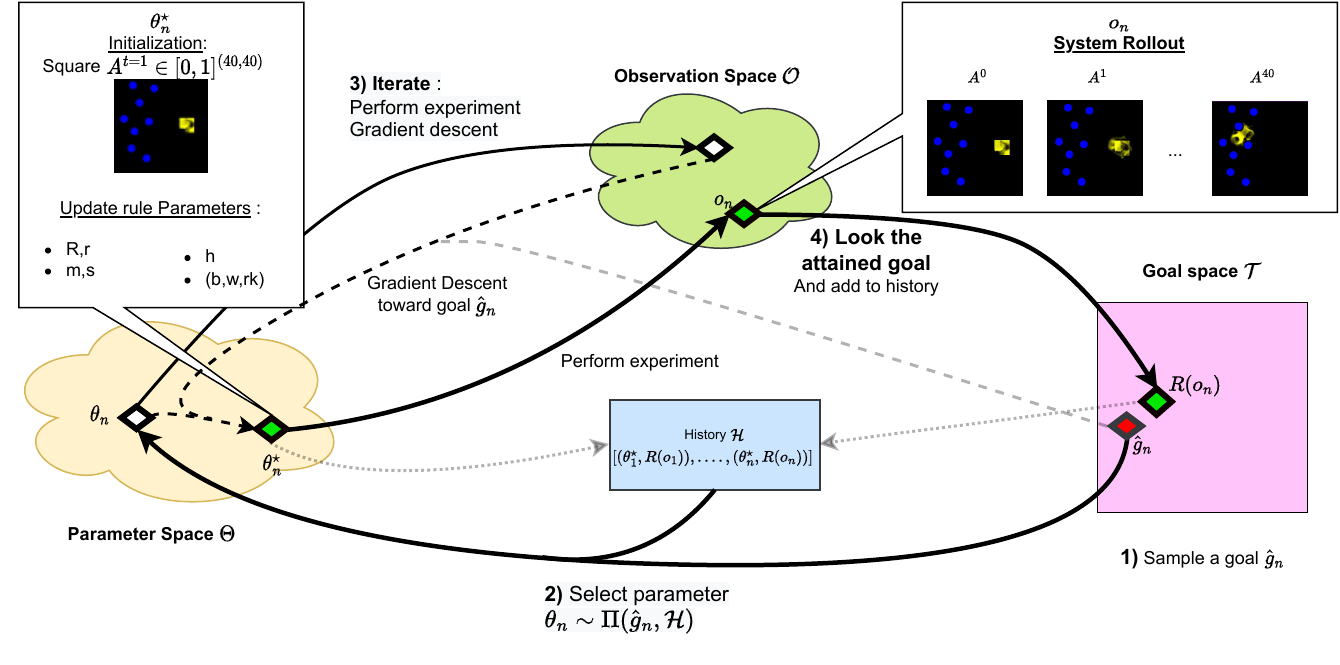}
\caption{IMGEP loop}
\label{appendix_fig:IMGEP}
\end{figure*}

\begin{algorithm}[!h]
\caption{IMGEP pseudo code}
\label{algo:ad_framework}
\vspace{0pt}
\begin{algorithmic}

\State \textbf{Initialization: } history $\mathcal{H}$ and models $\mathcal{T}, \Pi, Optim, R$.

\For{i=1..N}
\State Generate a target goal $\tau_i \sim \mathcal{T}(\mathcal{H})$ \Comment{use of \textit{curriculum learning} and \textit{diversity search}}

\State Train parameters on target goal
$\theta_i^* = Optim (\theta_i | \tau_i)$, where $\theta_i \sim \Pi(\mathcal{H} | \tau_i)$ \Comment{use of \textit{gradient descent} and \textit{stochasticity handling}}

\State Evaluate parameters $x_i \sim R(\theta_i^*)$ \Comment{\textit{behavioral characterization}}

\State Store in history $ H \gets H \cup (\theta_i^*, x_i) $ \Comment{\textit{reuse knowledge} for task sampling and training}

\EndFor
\Return $\mathcal{H}$
\end{algorithmic}
\end{algorithm}

In this section, we first recall the basics of the IMGEP procedure and then go into the details of each element of the method. 

Our method described in the pseudo code \ref{algo:ad_framework} starts by initializing a pool of (parameters, reached position) couples by random search, this constitutes the initial state of the history $\mathcal{H}$ (details in \ref{appendix:init_history}). Then, at each iteration, the method iterates through the following steps (illustrated in Fig.~\ref{appendix_fig:IMGEP}). \textbf{1) Sample a new goal} using a goal sampling strategy which takes into account the previously reached positions (details in \ref{appendix:goal_sampling}). An example of the sampling distribution can be found in green in Fig \ref{fig:curriculum}.a. \textbf{2) Infer starting parameters for that goal} by selecting parameters $\{(\theta_l,A_l^{t=1})_i\}_{i=1 ... t-1}$ associated to a previously reached position in history $\mathcal{H}$ that is close to the sampled goal (details in \ref{appendix:history_reuse_filter}). \textbf{3) Optimize parameters toward the sampled goal} by iteratively performing rollouts of the Lenia system under different environmental conditions $A_f$ and applying stochastic gradient descent on the MSE loss between the disk at goal position and the mass of the learnable channel at the last timestep (details in \ref{appendix:gradient}). \textbf{4) Update history $\mathcal{H}$} with the newly obtained parameter point and test it in various environmental conditions $A_f$ to estimate its reached position (details in \ref{appendix:param_eval}) (such that it can be later reused as a starting point for achieving other sampled goals).

As described in the main text, the behavioral space is the position (x,y) of the center of mass at the last timestep of the rollout. The loss we use is the Mean square error loss between the learnable channel at the last timesteps of the rollout and the same grid with a superposition of 2 disk centered at the goal position in the first channel. The target disk has this formula: $0.9x(0.15x(R_g<10)+0.85x(R_g<5))$ where $R_g$ is the euclidian distance to the goal position. 

To introduce more diversity in the search (and potentially getting out of difficult optimization landscape), some steps of IMGEP add mutation to the promising parameters before applying the optimization through gradient descent. More details can be found in \ref{appendix:mutation_details}.

Note that we also introduce an automatic way for the method to restart again from scratch in case of not good enough first steps (not present in pseudocode \ref{algo:ad_framework}). We refer to subsection.\ref{appendix:imgep_init_selection} for a detailed description of this restarting mechanism. 

Note that the goal positions as well as the measured reached positions (details in \ref{appendix:param_eval}) are normalized and centered between -0.5 and 0.5 (so that obstacle positions are at x > 0 ,Fig.\ref{fig:curriculum} in the main text) according to the map size $SX$. 

The following sections provide additional details about different parts of the method.

\subsubsection{Intialization of history}\label{appendix:init_history}

The IMGEP method first applies an initialization of history $\mathcal{H}$ through random search to bootstrap the whole IMGEP procedure. 

In this work, the initialization of history consist of 40 trials of random parameters. The range used for this random search are the one presented in \ref{appendix:lenia_params} except that we divide the strength of the kernels parameters h by 3. This change is done in order to have weaker/slower updates increasing the chance to have a pattern not exploding or vanishing in 50 timesteps, in order to facilitate further optimization. 

 This dividing of h by 3 is only to make things go faster (requiring less trials for the initialization of history) with some human heuristic on the system but should not be mandatory as random search without this should also get interesting parameters for initialization with more trials.  
 
\subsubsection{Warming up goal sampling}\label{appendix:warming_deterministic}

To accelerate the curriculum, we start the first 8 steps of the IMGEP with a deterministic goal sampling which tries to go as far as possible on the x axis. The goal position starts at position (-0.19,0) and is shifted of +0.06 along the x axis for every of those deterministic steps. The rest of the goal sampling is stochastic as described in \ref{appendix:goal_sampling}.

\subsubsection{Initialization selection}\label{appendix:imgep_init_selection}


History initialization and the first IMGEP steps have a huge impact on the performance of the method, as it will provide the basis for all subsequent optimization. History initialization and the warm up of goal sampling have a huge impact on the performance of the method, as it will provide the basis for all subsequent optimization.

To mitigate this problem, we also apply initialization selection with the objective of facilitating further optimization. We run the first steps of the method (random initialization and few steps of optimization), and observe the loss for the 3 first deterministic targets (described in section \ref{appendix:warming_deterministic}). If this loss is above a certain threshold for one of the 3 step, we start over again getting rid of the initialization history and initializing it again with random search. We perform this until we find a “good” initialization that is below the threshold for the 3 steps. 

\subsubsection{Goal sampling} \label{appendix:goal_sampling}

The goal sampling we chose in this work intends to sample goals ((x,y) positions) that should be most of the time further in the grid (for harder goals), not too far from previously reached positions (for feasibility of the goal) and also not too close from previously achieved goals (to make progress) . From those heuristic we introduce our engineered goal sampling strategy in pseudo code \ref{algo:goal_sampling}. The objective of this engineered sampling is to accelerate the search but much simpler ones could work if given enough computational budget (see ablation with totally random sampling \ref{appendix:imgep_randomsampling}).


\begin{algorithm}[!h]
\caption{Goal sampling strategy}
\label{algo:goal_sampling}
\vspace{0pt}
\begin{algorithmic}

\State \textbf{Input: } history $\mathcal{H}$ 
\State nb\_close=0,nb\_veryclose=0
\While{nb\_close<1 or nb\_veryclose >2:}
    \If{rand $\sim \mathcal{U}(0,1)$ <0.2 :} 
        \State goal =  bestgoal($\mathcal{H}$)+$( \mathcal{U}(0,1)\times0.04 + 0.02,(\mathcal{U}(0,1)\times0.45-0.22)/4,) $
        \Comment{Try little further than previous best}
    \Else
        \If{rand $\sim \mathcal{U}(0,1)$ <0.7 :} \Comment Try random far points 
            \State goal=$(-\mathcal{U}(0,1)\times0.2+0.35,- \mathcal{U}(0,1)\times0.45-0.22) $
        \Else
            \State goal=$(-\mathcal{U}(0,1)\times0.35+0.35,- \mathcal{U}(0,1)\times0.45-0.22) $
        \EndIf
    \EndIf
    \State nb\_close,nb\_veryclose=calc\_distances(goal,$\mathcal{H}$) 
\EndWhile
\Return goal
\end{algorithmic}
\end{algorithm}


\subsubsection{Mutation}\label{appendix:mutation_details}

We apply mutations on candidates parameters in order to increase diversity. Some mutations can facilitate optimization while others can lead to undesirable configurations impairing it. For this reason, we apply less gradient steps on those mutated parameters. See section~\ref{appendix:imgep_hyperparams} for the hyperparameters in this work.

In addition, we generate mutations of a parameter configuration until it results in a pattern not collapsing after 50 timesteps. For this (approximate) collapsing measure, we use a simple soft filter checking if the total mass in the learnable channel at the last timestep is >10 ( to test for death of matter) and if the mean square error between the learnable channel at the last timestep and the disk defined in \ref{appendix:imgep_details} centered on the center of mass of the learnable channel is < 25 (as a proxy for explosion of the mass, more details in \ref{appendix:param_eval}). This loop of mutations is counted in the total number of rollout performed by the IMGEP.

We refer to section \ref{appendix:lenia_mutation_range} for the mutation (distribution, mean, variance) applied to each parameters in the method.

\subsubsection{Gradient descent}\label{appendix:gradient}

Differentiating through Lenia can be difficult because the gradient must backpropagate through several steps (which moreover have their result clipped between 0 and 1) without vanishing. We should thus limit ourselves to a few iterations when training: in our experiments the loss is applied after 50 steps in Lenia.

Obtaining gradients that are informative for optimization requires an overlapping between the  mass in the learnable channel and the disk centered at the goal position. The curriculum we introduce in the goal sampling procedure (\ref{appendix:goal_sampling}) facilitates this overlap by generating goals that neither too far nor too close from the initial pattern at t=0 and from previously reached goal.

We refer the reader to appendix section A.\ref{appendix:imgep_nograd} for an ablation of the gradient descent showing the importance of it in the method.

\subsubsection{Parameter evaluation} \label{appendix:param_eval}

 We perform an evaluation of the parameters after each IMGEP step (sampling of goal and optimization of parameters). This evaluation consists of running 20 rollouts of 50 timesteps (the same rollout length as in the optimization rollout) with different random obstacle configurations and measures the average reached position over those rollouts.  

For each rollout, we also compute the mean square error between the learnable channel at the last timestep and the disk shape centered on the center of mass of the learnable channel at last timestep. We then take the average value over the rollouts. This is used as a proxy ``collapsing measure'' (explosion or death of the pattern) to apply a soft filter when selecting promising initialization parameter for a new goal as explained in section \ref{appendix:history_reuse_filter}. 

The parameters ($A_l,\theta_l$), the measured reached position $(r_x,r_y)$ and collapsing proxy measure $c$ are then stored in the history $\mathcal{H}$. 

\subsubsection{Reusing history $\mathcal{H}$ for a new goal.} \label{appendix:history_reuse_filter}

Once a goal is selected, we compute the L2 distance between all vectors $(c,r_x,r_y)$ of the history and $(c_{goal},g_x,g_y)$, where $g_x$ and $g_y$ are the (x,y) coordinate of the goal and $c_{goal}$ is a constant equal to 0.065 in this work. These L2 distances are used to select a point in the history reaching a position close to the goal while mitigating the risk of collapsing. 

In addition to these L2 distances for the selection of potential candidates for a new goal, we also filter out the points in the history having $c> 0.11$ allowing to remove the potential collapsing ones even though they might be close to the goal. We also take into account this collapsing proxy measure as collapsing parameters are hard to recover from through gradient descent.

The candidate parameter for a goal is therefore the point in the history which has $c<= 0.11$ and which minimize the L2 distance presented above.

\subsubsection{IMGEP search Hyperparameters}\label{appendix:imgep_hyperparams}

\begin{itemize}
    \item Number of IMGEP steps : 120
    \item History initialization : 40 trials of random parameters. 
    \item In 4 out of 5 IMGEP step, we mutate the candidate parameter before gradient descent. 
    \item Number of gradient steps : 125 when no mutation beforehand (1 out of 5 IMGEP steps)  , 15 when mutation beforehand.
    \item Rollout length : 50 timesteps
    \item Grid size : 256x256 
    \item Number of obstacle during the search: 8
    \item Initialization position on the 256x256 grid: [36:76,105:145]

\end{itemize}

\subsection{Basic obstacles tests and generalization tests} \label{appendix:tests}

Note that the tests we provide are proxy measure of agency/stability, and so what we present here are what we consider in this paper as agency. It is for example impossible to test for infinite time stability in finite time budget. Our stability tests are based on previous work on Lenia \cite{reinke2020intrinsically}.

\subsubsection{Empirical agency test}\label{appendix:agency_test}
 We describe here the agency test used in the paper:
 
 We first apply a prefilter to the obtained parameters by running a rollout of 500 steps with the obtained parameters. From this rollout, we measure if the mass at the last timestep was strictly above 0 (not dead) and below 6400 (explosion). The number are arbitrary and relatively ``loose'' so that we reject nearly no ``false positive''. This prefilter allows to throw out obvious non interesting parameters to reduce the computational cost of testing all obtained parameters -- especially for the random search method where many of them are not interesting. 

We then do rollout of 2000 timesteps for the empirical agency and moving test. The rollout is long (especially relative to the 50 timesteps of the search) in order to probe for long term stability. We compute some stats, from the rollout observations, which are used for the empirical agency test (and moving test) of the parameters inspired by \cite{reinke2020intrinsically}.

The empirical agency test consist of :

\begin{itemize}
    \item  Measuring if the mass of the learnable channel is > 0 and <6400 ($\sim$10\% of the map) at the last timestep of the rollout as those correspond to collapse and explosion.
    
    \item Measuring if the average mass is augmenting or decreasing too much between 2 windows of the rollout. This is a proxy measure for long term instability meaning that a big loss or increase of mass between the 2 windows is most of the time an indicator for long term instability. In this work, we measure the ratio between the average mass during the 0 to 500 window and 1500 to 2000 window. If this ratio is greater than 2, the parameters do not pass the test. The windows are relatively large to still allow for variation of mass during a rollout and the formation of a pattern in the first window.
    
     \item We also want the emerging pattern to be a spatially localized \textbf{Soliton} (ie pattern forming a single entity not expanding indefinitely, with a bounded radius). 
     To measure this, we perform a connectivity analysis of the pattern depending on the kernel radius, rejecting patterns where two distinct blobs of mass cannot influence each other (distance between blobs $\geq R*max(r)$).
     
\end{itemize}

\subsubsection{Moving test} \label{appendix:moving_test}

To test if a pattern passing the empirical agency test is moving, we measure if the center of masse of the learnable channel moved further than 100 pixels from the initialization position at any point during the 1000 first steps of the rollout. 

\subsubsection{Speed measure}\label{appendix:speed_measure}

To measure speed of agents, we use the 2000 timesteps rollout computed in the filter phase and track the average distance travelled by the center of mass of the agent on sliding overlapping windows of size 25 starting from timestep 150 to timestep 2000. The result is divided by 25 (the size of the sliding window) in order to have a per timestep average distance travelled.  We use a sliding window to filter slight back and forth movement of the center of mass (which can even be due to self organization without clear ``movement'' of the whole). 
Note that we compute the speed only for agents passing the filters above.

The same is done to measure speed with obstacles but we average on the 50 rollouts with random obstacles computed in the robustness test. The only small modification is that if an agent does not pass the survival tests above on the rollout (for example its mass reaches 0 ), we set the speed for this rollout to 0.  

\subsubsection{Basic obstacles tests}\label{appendix:basic_osbtacle_test}

We then test the parameters leading to moving agents by performing 50 rollouts of 2000 timesteps where obstacles are the same as in training i.e. obstacles of radius 10. We place 24 obstacles in the whole grid (compared to only the right part of the grid in training), from which 23 are randomly placed and one being in the trajectory of the moving agent to be sure that it will encounter at least one obstacle in the rollout. To do this we look at the achieved position of the moving agent without obstacle at timestep 1000 and put an obstacle here in the test for every rollout. We also remove any obstacle pixel in the initialization area (pixel of the learnable channel >0 at the initialization) as well as in a radius of 10 pixels (euclidian distance) of the initialization (to let some space for the initialization to develop). 

From the observations of the rollout we compute the same statistics and same categories used for the agency test. To get the robustness measure we then measure the fraction of rollout where the pattern pass the empirical agency test.

\subsubsection{Generalization tests}\label{appendix:generalization_tests}

Here is a full description of each of the generalization test conducted in the \textit{Generalization} section in the main text. For all the quantitative generalization tests, we used the same robustness test as above except that we do it on 10 random trials instead of 50: we run rollout of 2000 timesteps, then measure if it fulfills the empirical agency test. The measure of robustness is again measured by the proportion of trials where the agent pass the empirical agency test. (hence between 0 and 1). 

We also provide a more detailed table of generalization results in Tab.\ref{appendix:generalization_table} adding also agents obtained through random search and semi-manual search.

\begin{itemize}
    \item \textbf{Initialization noise}. In this experiment, we add a centered gaussian noise to the pixel of the initialization square $A^1$. In the first test ``init noise rate'' we vary the proportion of pixels affected by this gaussian noise, testing proportions [0.2,0.4,0.6,0.8,1.], and keep the variance fixed to 1. In the ``init noise std'' test, we apply the noise to all pixels of the initialization but vary the variance of the gaussian in [0.5,1.5,2.5,3.5,4.5].   
    
    \item \textbf{Obstacles} In all of these test we also remove obstacles pixel from the initialization square and in a radius of 10 pixels (euclidian distance) around it.
        \begin{itemize}
        \item \textbf{Obstacle radius} In this test, we vary the radius of the obstacles in [4,7,10,13,16]. The number of obstacles varies according to the radius of obstacles to keep the same ratio of obstacle pixels with the default one which is 24 obstacles of radius 10. The formula is $\text{Number obstacles}=24 \times(10/var)^2$.
        \item \textbf{Obstacles number} In this test, we vary the obstacle number keeping the radius fixed to the default one (radius=10). We try obstacle number= [24, 30,36,42 ,48] . 
        \item \textbf{Obstacle speed}. In this test, we change the dynamic of the obstacle channel so that obstacle move at a certain speed as detailed in \ref{appendix:lenia_obstacles}. For a speed of 1, the obstacle channel is shifted of 1 on the left at every timestep, for a speed of 0.5, the obstacle channel is shifted of 1 every 2 timesteps. We tested obstacle speed of [1/3,1/2,1,2,3]. In this test we put 24 obstacles of radius 10. 
        
        \end{itemize}
        
    \item \textbf{Scale} In this test, we vary the scale of agents by changing their kernel size multiplying the parameter $R$ of the simulation by the factor. A smaller (resp bigger) size of kernel means that the convolution will cover a smaller (resp bigger) neighbourhood. We also change the initialization size by a factor $\alpha$ to match the scale. To do this, we use a downscaling (or upscaling) of the initialization $40 \times 40$ square with bilinear interpolation. We test both smaller sizes : 0.15,0.65 , as well as bigger sizes: 1.15,1.65,2.15.  
    
    \item \textbf{Update}. In this tests, we perturb the update (what is added to the current state) from step 0 until step 1900. We let the step from 1900 to 2000 free of update perturbation to allow the rule to recover until step 2000 for the statistics computation. 
        \begin{itemize}
        \item \textbf{Update mask} In this test, for a value of update mask p<1, every pixel has a probability p of being updated while the rest of the pixels will keep the same value. This does not apply to the update applied by the obstacles. For a value 1<p<2, each pixel is updated one time using the update rule normally (sensing and add of growth) giving a new state and then each pixel is updated again from this new state with a $p-1$ probability (the sensing on the potential second random update is done by sensing the new state). We test the update mask rate in [0.2,0.6,1.,1.4,1.8].
        \item \textbf{update noise std} In this test, we add noise to the update of the learnable channel before the clipping as such :  $$A^{t+1}_l=A^t_l + \frac{1}{T} \left(G(K*A^t)+\mathcal{N}(0_{256\times256},\sigma \mathbb{I}_{256\times256})\right)$$ where $\mathcal{N}(\mu,\Sigma)$ is a gaussian vector of mean $\mu$ and variance $\Sigma$. We vary $\sigma$ in $[0.5,1.5,2.5,3.5,4.5]$
        \item \textbf{Update noise rate}. We add noise to the update of the learnable channel before clipping. Every pixel has a probability $p \in [0.2,0.4,0.6,0.8,1.]$ to have a gaussian noise of mean 0 and variance 1.  
        \end{itemize}

    \item \textbf{Morphological computation/ Hand damage}. In this test, we allow an exterior experimenter to pause the simulation and put pixels of the learnable channel to 0. After the damage, we then let the simulation unroll as usual starting from the damaged state $A_l^{damaged}$.
    \item \textbf{Interactions (Multi agents setting)}. We allow to put several initialization square in the learnable channel. As the update rule apply to all the grid the same way, if a couple (initialization square, update rule) already led to a an agent in the case of a single initialization square then several of them that are not interfering ( further enough so that the convolution of a pixel of one does not contains pixels of the other) will lead to several agents. 

    \item \textbf{Custom obstacles}. We allow an experimenter to freely draw obstacle in the grid. This allows to have obstacles with shapes not seen during training. 
    \item \textbf{Custom init states} In this test, we replace the initialization of the pattern (that was optimized) by simple arbitrary shape such as disk with a gradient (the gradient being to have an asymmetry for movement), disk of large size etc. The web demo at \href{http://developmentalsystems.org/sensorimotor-lenia-companion}{http://developmentalsystems.org/sensorimotor-lenia-companion} also allows to load any image as initialization of the system.

    \item \textbf{External control}\label{appendix:attraction} This experiment consists in adding a new channel (a new type of cell) to the system which we want to act as an attractive element. We conducted a semi handmade search in order to search for a rule, sensing in the attractive channel and updating the learnable channel, leading to this attractive behavior. 

    Note that this attractive element should attract but not disturb too much the matter as we don't want the attractive matter to be able to destroy the agent dynamics. 
    
    In fact, we first searched for a rule tuned for one agent found with the IMGEP search (ie one parameter point ($A_l,\theta_l$)). By doing so, the rule is adapted to the dynamic of this specific agent (for example different agents might have different range for pixel value or growth etc). 
    
    The search for a rule (tuned for a specific agent) is semi handmade. We first preselect some rule parameters from a set of random rules. The preselection is done by moving a circle of attractive mass along a predefined straight trajectory in an environment with a moving agent. We then look if the attractive mass and the agent overlaps at the last timestep which should mean that the agent followed this attractive mass. An experimenter then select by hand the rules that lead to attraction of mass without too much perturbation by controlling the mass of attractive matter in a real time simulation with the moving agent. 
    
    After searching for a rule for a specific agent, we then tested it on some other moving agents obtained with IMGEP. Some agents (some parameters ($A_f,\theta_f$)) are more prone to work with it (meaning attraction while not affecting the stability too much) while it destroy the stability of others. The reported qualitative results on this test are performed on agents where the rule leads to stable attraction.
            
\end{itemize}

\subsection{Comparison baselines}\label{appendix:comparison_baselines}

\subsubsection{Random search details} \label{appendix:random_search}

We use uniform sampling of parameters with the ranges given in \ref{appendix:lenia_params}. 

The initialization 40x40 square is randomly sampled with each of the pixel constituting it being independently sampled following a uniform distribution between 0 and 1. 


\subsubsection{``Handmade'' agents (from original Lenia paper)}\label{appendix:handmade}

The parameters from this dataset are the one from the original Lenia paper \cite{chan2019lenia,chan2020lenia} (following these links: \href{https://github.com/Chakazul/Lenia/tree/master/Python/found}{\tiny{https://github.com/Chakazul/Lenia/tree/master/Python/found}}, and \href{https://github.com/Chakazul/Lenia/blob/master/Python/old/animals.json}{\tiny{https://github.com/Chakazul/Lenia/blob/master/Python/old/animals.json}}). Contrary to the rest of the paper we use the classic parameterization of Lenia for the agent channel. We filter out those that have more than one channel or an initialization that has a side bigger than 256. We then apply the pre-filter and filter as explained in section \ref{appendix:agency_test}. We provide the resulting parameters in the data folder of \href{https://github.com/flowersteam/sensorimotor-lenia-search}{\tiny{https://github.com/flowersteam/sensorimotor-lenia-search}}. 

In the handmade search from the original Lenia papers, self-organizing patterns were discovered by basic evolutionary algorithms, through one of these routes: (1) random parameter values and initial patterns; (2) start from an existing moving pattern and mutate the parameter values; (3) manual editing of the initial pattern.

\subsection{Movie legends}\label{appendix:legend_videos}

You can find all movies on this companion website \href{https://developmentalsystems.org/sensorimotor-lenia-companion/}{\tiny{https://developmentalsystems.org/sensorimotor-lenia-companion/}}.

\begin{itemize}
    \item \textbf{Movie S1: Sensorimotor agents} Different agents (yellow) emerging from rules obtained by the IMGEP. The agents display sensorimotor capabilities: they are robust and react to perturbations by the obstacles (blue). The righmost video shows the system with a different colormap (fixed obstacle channel in black) to highlight the differences in activity in the agent as a response to perturbation. 
    \item \textbf{Movie S2: Random search } Each 100 squares are random parameters trials (each 1 channel and 10 rules so \~ 130 parameters for all the rules of a square). We observe that a lot of random search trials lead to death or explosion of the mass. Very little lead to stable spatially localized pattern and even less to moving ones. 

    \item \textbf{Movie S3 Orbium, moving agent from the original lenia papers, fragile to external perturbations} S3.a: Orbium: the equivalent of the glider in Lenia (from the original lenia paper), an example of moving agent. S3.b and S3.c videos: collision between several orbium leading to death/explosion. This shows the fragility of the orbium to external perturbations. 

    \item \textbf{Movie S4 Orbium perturbed by obstacles} Orbium, equivalent of the glider in Lenia (from the original lenia paper), dies from perturbations by obstacles. 

    \item \textbf{Movie S5: Agents obtained by each method} 100 Patterns passing our agency tests obtained by each method: random search(S5.a),IMGEP (S5.b), handmade search ((S5.c)from Lenia original papers). A lot of IMGEP obtained agents are moving agents with high speed while a lot of agents obtained by random search are static. 
    
    \item \textbf{Movie S6: Moving obstacle test on agents obtained by each method} 100 Agents obtained by random search(S6.a),IMGEP(S6.b), handmade search ((S6.c) from Lenia original papers). We observe that the proportion of agents with robustness to moving obstacles is much higher in the agents obtained by IMGEP than the ones obtained by random search and handmade search. 

    \item \textbf{Movie S7: Illustration of the quantitative generalization tests performed} See companion website \href{https://developmentalsystems.org/sensorimotor-lenia-companion/}{\tiny{https://developmentalsystems.org/sensorimotor-lenia-companion/}}. Videos of quantitative tests for 2 moving agents obtained by IMGEP. We display only a subset of the value tested for every quantitative test. 

    \item \textbf{Movie S8: Out of distribution obstacles: Different shapes} Test of a moving agent obtained by IMGEP on obstacles that were not seen during training. 

    \item \textbf{Movie S9: Out of distribution obstacles: maze.} Test of a moving agent to maze like obstacles. 

    \item \textbf{Movie S10: Out of distribution obstacles: Bullet like obstacles} Test of a moving agent to bullet like environment: fast small moving obstacles. 

    \item \textbf{Movie S11: Individuality preservation} Example of moving agents obtained by IMGEP colliding while keeping their individuality, they don't merge or collapse from the collision. 

    \item \textbf{Movie S12: Reproduction} For some moving agents, under specific conditions, the collision of 2 agents can lead to the self-organization of a 3rd agent. (each with its own individuality) 
    \item \textbf{Movie S13: Attraction} Example of moving agents attracting each other while still maintaining their own individuality. 
    \item \textbf{Movie S14: Asynchronous update} Testing a moving agent with asynchronous updates. Each cell is updated with a certain probability at each step leading to cells being asynchronously updated. 
    \item \textbf{Movie S15: Scaling the agents down} The moving agents size is reduced. The scaled down agents still seem to behave similarly (same shape and have sensorimotor capabilities) to the normal size one while being composed of less cells. 
    \item \textbf{Movie S16: Morphological computation} We pause the simulation and remove some cells of a moving agent. As a response to this alteration of the structure, the moving agent changes direction, regrow itself and moves away. This video isolates the fact that the macro agent senses perturbations of its structure and respond to it by a morphological growth. 
    \item \textbf{Movie S17: External control.} We introduce an attractive element in another channel (in Cyan). We learned the rule that control the way this external element channel influences the learnable channel(Yellow) and display the resulting behavior here. The moving agent is effectively attracted to this introduce component. By controlling the external element we can control live the direction of the moving agent. 
    \item \textbf{Movie S18: Robustness to initialization} Testing the robustness of the learned rule to emerge an agent from different initial patterns. We replace the learned initial pattern by : S18.a a disk with a gradient; S18.b a large disk (much larger than an agent); S18.c top a disk with gradient of another size, bottom a disk without gradient. Some initialization lead to the robust emergence of one or several agents while some lead to the collapse of the pattern. 
    
    \item \textbf{Movie S19: Examples of agents considered non moving by our moving test}
\end{itemize}

\end{document}